\documentclass[twocolumn,showpacs,preprintnumbers,nofootinbib,prd,superscriptaddress,groupedaddress,10pt]{revtex4-1}

\usepackage{graphicx,amssymb,amsmath,amsthm,amsfonts,epsfig}
\usepackage{bm}
\usepackage{dcolumn}
\usepackage[latin1]{inputenc}
\usepackage{latexsym}
\usepackage{rotating}
\usepackage{longtable}
\usepackage{enumerate}
\usepackage{mathrsfs}
\usepackage{mathtools}
\usepackage{url}
\renewcommand{\vec}[1]{\boldsymbol{#1}}

\def\be{\begin{equation}}
\def\ee{\end{equation}}
\def\bea{\begin{eqnarray}}
\def\eea{\end{eqnarray}}
\newcommand{\beq}{\begin{eqnarray}}
\newcommand{\eeq}{\end{eqnarray}} 
\newcommand{\ba}{\begin{align}}
\newcommand{\ea}{\end{align}}

\def\ba{\bar{a}}

\begin{document}
{\hfill }

\author{
Taishi Ikeda$^{1}$}
\email{taishi.ikeda@tecnico.ulisboa.pt}
\author{Tomohiro Nakamura$^{2}$}
\email{nakamura.tomohiro@g.mbox.nagoya-u.ac.jp}
\author{Masato Minamitsuji$^{1}$}
\email{masato.minamitsuji@ist.utl.pt}
\affiliation{${^1}$ CENTRA, Departamento de F\'{\i}sica, Instituto Superior T\'ecnico -- IST, Universidade de Lisboa -- UL, Avenida Rovisco Pais 1, 1049 Lisboa, Portugal\\
${^2}$Division of Particle and Astrophysical Science, Graduate School of Science, Nagoya University, Nagoya 464-8602, Japan
}

\newcommand{\com}[1]{\textcolor{blue}{\\\sf{[Comment out: #1]}} }
\title{Spontaneous scalarization of charged black holes in the Scalar-Vector-Tensor theory}
\begin{abstract}
We present spontaneous scalarization of charged black holes (BHs) which is induced by the coupling of the scalar field to the electromagnetic field strength and the double-dual Riemann tensor $L^{\mu\nu\alpha\beta}F_{\mu\nu}F_{\alpha\beta}$ in a scalar-vector-tensor theory.
In our model, the scalarization can be realized under the curved background with a non-trivial electromagnetic field, such as Reissner-Nordstr$\ddot{\rm o}$m Black Holes (RN BHs).
Firstly, we investigate the stability of the constant scalar field around RN BHs in the model, and show that the scalar field can suffer a tachyonic instability.
Secondly, the bound state solution of the test scalar field around a RN BH and its stability are discussed.
Finally, we construct scalarized BH solutions, and investigate their stability.
\end{abstract}
\maketitle
\section{Introduction}
Black holes (BHs) are the most important astrophysical objects predicted by general relativity(GR) and modified gravity theories.
\
In Einstein-Maxwell theory, due to the no-hair theorem\cite{Bekenstein:1971hc,Bekenstein:1995un}, BHs have only three parameters, which are its mass, angular momentum, and charge.
It means that a BH in GR is the simplest compact object.
\footnote{
We can construct BHs with scalar hair in GR with complex massive scalar fields \cite{Herdeiro:2014goa}, 
and BHs with Proca hair in GR with complex Proca fields \cite{Herdeiro:2016tmi}.
}
On the other hand, to solve the dark energy or the dark matter problem,
several gravitational theories beyond GR have been advocated.
They introduce additional degrees of freedom as the dark side of our Universe.
In some modified gravity theories, BHs can have additional parameters corresponding to the additional fields.
Such BHs are called hairy BHs.
Since several properties of hairy BHs are different from those of BHs in GR,
observations using gravitational waves\cite{Abbott:2016blz}, BH shadows\cite{Akiyama:2019cqa}, and motion of galaxies\cite{Hui:2012jb} can give us constraints on the modified gravity.
Therefore, it is important to investigate which modified gravity theories possess hairy BHs, or which theories satisfy the BH no-hair theorem.
\par
A modified gravity theory which has an additional scalar field is called a scalar-tensor (ST) theory, and a theory which has a scalar field and a vector field is called a scalar-vector-tensor (SVT) theory.
Recently, the most general framework of the SVT theories with second-order equations of motion was constructed\cite{Heisenberg:2018acv} (see review paper \cite{Heisenberg:2018vsk}).
This theory includes not only ST theories but also several modified gravity theories which have a vector field.
Focusing on the case of the vanishing vector field, the theory reduces to the Horndeski theory\cite{Horndeski:1974wa,Kobayashi:2011nu,Kobayashi:2019hrl}, which is the most general ST theory without Ostrogradsky ghosts\cite{Woodard:2015zca}.
The SVT theory is divided into two classes.
The first class is the theory with the $U(1)$ symmetry, and the other one is the theory without the $U(1)$ symmetry.
Since the $U(1)$ symmetry constrains interactions and solutions, in this paper, we consider the theory with the $U(1)$ symmetry.
We will summarize the SVT theory with the $U(1)$ symmetry in Appendix \ref{sec:Scalar-vector-tensor theories with U(1) symmetry}.
\par
In the Horndeski theory with the shift symmetry of the scalar field, the symmetry strongly constrains BH solutions.
Under some conditions, the no-hair theorem of the shift-symmetric Horndeski theory was proven\cite{Hui:2012qt}(see also \cite{Babichev:2016rlq}).
In the theorem, the regularity of the Noether current $J_{\Phi}^{\mu}$ associated with the shift symmetry plays an important role.
We can naturally extend the no-hair theorem to the SVT theory with the U(1) and shift symmetries, and without a cosmological constant.
The conditions of the theorem for the SVT theory are as follows:
\begin{enumerate}
\item The spacetime is spherically symmetric and static spacetime with the asymptotic flatness.
\item $F$, $\tilde{F}$ and the gradient of the scalar field vanish in the asymptotic region.
\item The scalar and vector fields have the same symmetries with the metric.
\item The norm of the Noether current $J_{\Phi\mu}J^{\mu}_{\Phi}$ is finite on and outside the BH horizon.
\item The theory has the canonical kinetic term of the scalar field $X\subset f_{2}(X,F,\tilde{F},Y)$.
\item The functions $G_{i=3,4,5}(X)$, $f_{i=3,4}(X)$ and $\tilde{f}_{3}(X)$ are analytic at $X=0$.
\item $f_{2}(X,F,\tilde{F},Y)$ is analytic at $X=0$, $Y=0$, $F=0$, and $\tilde{F}=0$.
\item $f_{3}(X)$ vanishes at $X=0$.
\end{enumerate}
Here, $G_{i}$ and $f_{i}$ are functions of the SVT theory, and $F=F^{\mu\nu}F_{\mu\nu}$, $\tilde{F}=\tilde{F}^{\mu\nu}F_{\mu\nu}$ (see Appendix\ref{sec:Scalar-vector-tensor theories with U(1) symmetry}).
$F_{\mu\nu}$ and $\tilde{F}_{\mu\nu}$ are the field strength of the vector field, and its dual field (i.e. $F_{\mu\nu}=\nabla_{\mu}A_{\nu}-\nabla_{\nu}A_{\mu}$, $\tilde{F}^{\mu\nu}=\frac{1}{2}\epsilon^{\mu\nu\rho\sigma}F_{\rho\sigma}$).
$X$ is a kinetic term of the scalar field $\Phi$ (i.e. $X=-\frac{1}{2}\nabla^{\mu}\Phi\nabla_{\mu}\Phi$). 
Under these assumptions, the scalar field around BHs must be constant (see Appendix.\ref{Sec:No hair theorem for subclass with shift symmetry} for proof).
The class of the SVT theory which satisfies the above conditions does not possess BH solutions with a scalar hair.
On the other hand, BH solutions in the class which does not satisfy the conditions may have a scalar hair.
As an example, hairy BH solutions were constructed in the theory in which $f_{3}(0)$ does not vanish.\cite{Heisenberg:2018vti}.
The scalar field in the hairy BHs decays fast in the far region.
Thus, it may be difficult to distinguish these hairy BHs from BHs in GR at the asymptotic region.
The other possibility in which hairy BHs can be realized is violation of the shift symmetry.
Although it is expected that there are many types of hairy BH solutions in the theory, we focus on the hairy BH solutions associated with spontaneous scalarization.
\par
Spontaneous scalarization is the mechanism through which a relativistic compact object (either a BH or a star) in GR becomes unstable in the strong gravity regime, 
and spontaneously obtains a nonzero scalar hair with a nontrivial profile of the scalar field. 
Following the pioneering work by Damour and Esposito-Farese \cite{Damour:1993hw,Damour:1996ke}
spontaneous scalarization has been studied mainly in the context of a relativistic star, where an instability of a relativistic star in GR is caused by a tachyonic conformal coupling to matter. 
Because of the existence of the threshold value for the tachyonic coupling \cite{Harada:1998ge}, which is marginally consistent with constraints from binary-pulsar observations \cite{Freire:2012mg} (see \cite{Berti:2015itd} for a review), 
spontaneous scalarization occurs only in a highly compact star, 
and hence the difference from GR appears only in the strong gravity regimes.
\par
On the other hand, models for spontaneous scalarization of BHs have been proposed and studied very recently. 
The first models of BH scalarization were based on the Einstein-scalar-Gauss-Bonnet theory\cite{Silva:2017uqg,Antoniou:2017hxj,Doneva:2017bvd,Antoniou:2017acq,Minamitsuji:2018xde,Silva:2018qhn}. 
In these models, Schwarzschild solutions with the vanishing scalar field can be solutions if the coupling between the Gauss-Bonnet term and the scalar field satisfies the certain condition, and the scalar field suffers from the tachyonic instability in the vicinity of the BH horizon if the coupling constant is larger than the threshold value, 
and it is then expected that scalarized BH solutions are realized as the final state of the instability. 
The scalarized rotating BH solutions were also constructed in the same models\cite{Cunha:2019dwb}.
It was also discussed which class of ST theories can realize the spontenious scalarization in the context of the Horndeski theory \cite{Minamitsuji:2019iwp}.
\par
It was also discussed whether models of spontaneous scalarization of BHs can be successfully embedded into realistic cosmology, in the case that the scalar field has a cosmological origin.
The authors of \cite{Anson:2019uto} applied the scalar-Gauss-Bonnet model of scalarization to inflation, and
mentioned that a catastrophic instability develops and inflation is spoiled.
The authors of \cite{Franchini:2019npi} discussed the possibility that the scalar energy is sub-dominated in Universe,
and concluded that tuning of cosmological initial data is needed in order to keep the scalar field subdominant during cosmic evolution.
\par
Spontaneous scalarization can be realized in other models.
The simple example is SVT theories with coupling $f(\Phi)F^{\mu\nu}F_{\mu\nu}$ or $f(\Phi)\tilde{F}^{\mu\nu}F_{\mu\nu}$\cite{Herdeiro:2018wub,Fernandes:2019rez,Fernandes:2019kmh}.
In this model, scalarization can occur for the Reissner-Nordstr$\ddot{o}$m (RN) BH which satisfies a certain condition.
Scalarization of the ST theory coupled to Born-Infeld non-linear electromagnetic field was also discussed\cite{Stefanov:2007eq}.
\par
In this paper, we discuss the possibility of BH scalarization induced by the other type of a coupling in the SVT theory.
From the SVT theory without an Ostrogradsky ghost, the invariant quantities of the vector field which can induce the tachyonic instability of the scalar field are $F^{\mu\nu}F_{\mu\nu}$, $F_{\mu\nu}{}\tilde{F}^{\mu\nu}$, and $L^{\mu\nu\alpha\beta}F_{\mu\nu}F_{\alpha\beta}$ where
$L^{\mu\nu\alpha\beta}=\frac{1}{4}\epsilon^{\mu\nu\rho\sigma}\epsilon^{\alpha\beta\gamma\delta}R_{\rho\sigma\gamma\delta}$ is double-dual Reimann tensor
\footnote{$\epsilon^{\alpha\beta\gamma\delta}=\frac{1}{\sqrt{-g}}E^{\alpha\beta\gamma\delta}$
is covariant anti-symmetric tensor
where $E^{\alpha\beta\gamma\delta}$ is the Levi-Civita tensor with $E^{0123}=1$.}.
It was shown that the first two invariants can induce scalarization\cite{Herdeiro:2018wub,Fernandes:2019rez}.
In this paper, we discuss the possibility of scalarization which is induced by $L^{\mu\nu\alpha\beta}F_{\mu\nu}F_{\alpha\beta}$, 
As a simple toy model, we consider the following action:
\begin{eqnarray}
S&=&\int d^{4}x\sqrt{-g}\Bigl(
\frac{1}{16\pi G}R-\frac{1}{2}\nabla^{\mu}\Phi\nabla_{\mu}\Phi
-\frac{1}{4}F^{\mu\nu}F_{\mu\nu}
\nonumber\\
&&
+H(\Phi)L^{\mu\nu\alpha\beta}F_{\mu\nu}F_{\alpha\beta}
\Bigr),
\label{eq:action of f4 model}
\end{eqnarray}
where $g_{\mu\nu}$ is the spacetime metric, $R$ is the Ricci scalar associated with $g_{\mu\nu}$, 
$g$ is the determinant of the metric, and $\nabla_{\mu}$ is the covariant derivative with respect to $g_{\mu\nu}$.
$\Phi$ is the scalar field.
$H(\Phi)$ is a function of $\Phi$ which satisfies
\begin{eqnarray}
H(0)&=&0,\\
H'(0)&=&0,
\end{eqnarray}
where $H'(\Phi)=\partial_{\Phi}H(\Phi)$.
It characterizes the non-linearity of the scalar field.
The condition for $H(\Phi)$ ensures that the RN BH with vanishing scalar field is a solution, and scalarization of a RN BH is realized.
Since the model is subclass of the SVT theory, an Ostrogradsky ghost does not appear.
The coupling in the action Eq.\eqref{eq:action of f4 model} originates from the function $f_{4}(\Phi,X)$ in Eq.\eqref{action SVT}, and we call the model the $f_{4}$ model.
The SVT theory for spontaneous scalarization of charged BHs with analytic model functions has to be the model which has the coupling $H(\Phi)h(F^{\mu\nu}F_{\mu\nu},F^{\mu\nu}\tilde{F}_{\mu\nu})$, where $h$ is function of $F^{\mu\nu}F_{\mu\nu}$ and $F^{\mu\nu}\tilde{F}_{\mu\nu}$\cite{Herdeiro:2018wub,Fernandes:2019rez,Stefanov:2007eq}, or the model which has the coupling $H(\Phi)L^{\mu\nu\alpha\beta}F_{\mu\nu}F_{\alpha\beta}$ in the $f_{4}$ model.
\par
This paper is organized as follows:
In section II, we investigate the condition in which the tachyonic instability of the scalar field around RN BH is realized.
In section III, using the test field analysis, bound states around RN BHs are constructed, and the stability is investigated.
In section IV, we construct the scalarized BH solutions in the $f_{4}$ model.
\par
In this paper, we use units $(16\pi G=c=1)$ and mostly plus signature of the metric.
\section{Instability around RN BHs with vanishing scalar field}
\subsection{No hair theorem in the $f_{4}$ model}
The authors of \cite{Silva:2017uqg} considered the scalar-tensor theory with the coupling between the scalar field and the Gauss-Bonnet term, and discussed 
the condition of the BH no-hair theorem for the coupling.
If the condition is satisfied, the BH solution must have a trivial profile of the scalar field.
In this subsection, we apply the same argument to the $f_{4}$ model Eq.\eqref{eq:action of f4 model}, and 
discuss the no-hair condition.
\par
\begin{figure}[t]
  \centering
  \includegraphics[width=0.30\textwidth]{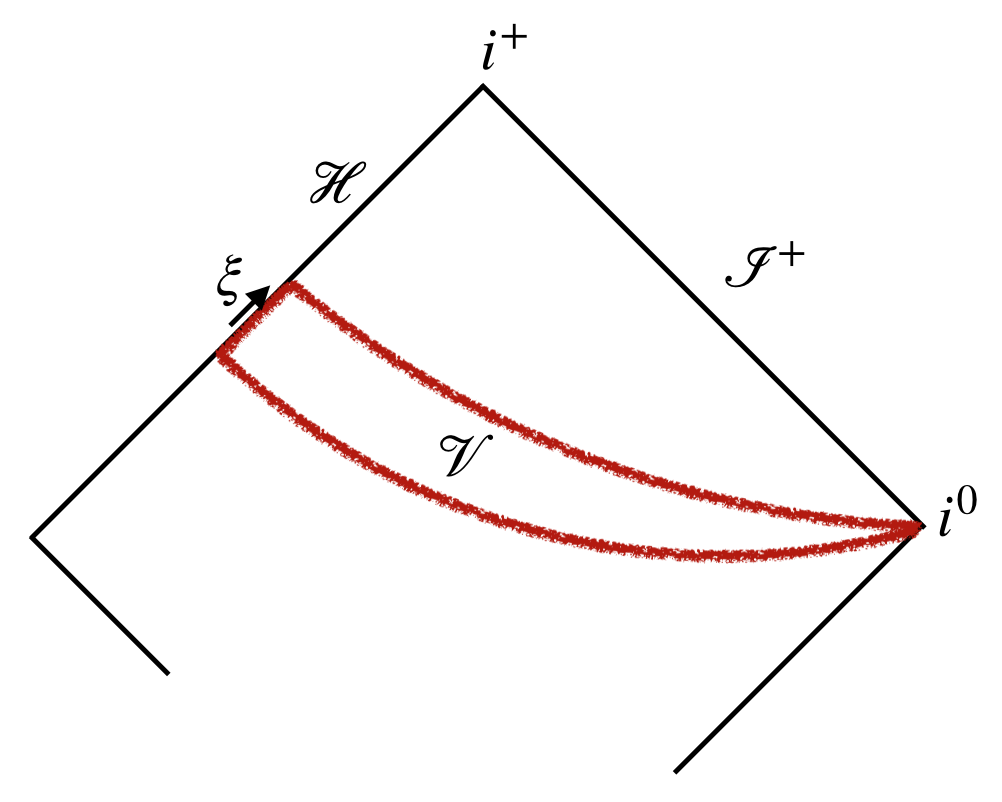}
  \caption{Definiton of $\mathcal{V}$.
  $\mathscr{H}$, $\mathscr{I}^{+}$, $i^{+}$, and $i^{0}$ are BH horizon, null infinity, time-like infinity, and spatial infinity, respectively.}
  \label{Fig:definition of V}
\end{figure}

We focus on the BH spacetime which is stationary and asymptotically flat.
From the symmetry, the spacetime has a time-like Killing vector $\xi^{\mu}$.
We assume that the scalar field is generated from the Killing vector : $\mathcal{L}_{\xi}\Phi=0$, and
the event horizon is a Killing horizon associated with the Killing vector.
We consider the region $\mathcal{V}$ which is enclosed by two Cauchy surfaces, a BH horizon and the spatial infinity (see Fig.\ref{Fig:definition of V}).
Varying the action with respect to the scalar field yields
\begin{eqnarray}
\nabla^{\mu}\nabla_{\mu}\Phi+H'(\Phi)L^{\mu\nu\alpha\beta}F_{\mu\nu}F_{\alpha\beta}=0.
\label{Eq:EOM scalar field}
\end{eqnarray}
Let us consider the following integration:
\begin{eqnarray}
\int_{\mathcal{V}}d^{4}x\sqrt{-g}
\left(
H'(\Phi)\nabla^{2}\Phi
+H'(\Phi)^{2}L^{\mu\nu\alpha\beta}F_{\mu\nu}F_{\alpha\beta}
\right)=0.\nonumber\\
\end{eqnarray}
Using the divergence theorem, we have
\begin{eqnarray}
&&\int_{\mathcal{V}}d^{4}x\sqrt{-g}\left(
-H''(\Phi)\nabla^{\mu}\Phi\nabla_{\mu}\Phi
+H'(\Phi)^{2}L^{\mu\nu\alpha\beta}F_{\mu\nu}F_{\alpha\beta}
\right)\nonumber\\
&=&-\int_{\partial\mathcal{V}}d^{3}x\sqrt{h}n^{\mu} H'(\Phi)\nabla_{\mu}\Phi,
\label{eq:integration}
\end{eqnarray}
where $H''(\Phi)=\partial_{\Phi}^{2}H(\Phi)$.
From the spacetime symmetry and the asymptotic flatness, the right hand side must vanish.
Outside the horizon, $\nabla^{\mu}\Phi\nabla_{\mu}\Phi$ is positive.
Therefore, if $H''(\Phi)L^{\mu\nu\alpha\beta}F_{\mu\nu}F_{\alpha\beta}<0$ inside $\mathcal{V}$, 
each term in the integrand of the left hand side in Eq.\eqref{eq:integration} must vanish, and the scalar field has to be constant.
\subsection{Tachyonic instability of scalar field arounf RN BHs in the $f_{4}$ model}
As is discussed in the previous subsection, the sign of $L^{\mu\nu\alpha\beta}F_{\mu\nu}F_{\alpha\beta}$ is important for a non-trivial scalar field profile.
In this subsection, we treat the scalar field as a test field, and consider the stability of the scalar field around 
the RN BH solution with the vanishing scalar field.
\par
The RN BH solution is a spherically symmetric charged BH, whose metric and vector field are given as
\begin{eqnarray}
ds^{2}&=&-f_{(\rm RN)}(r)dt^{2}+\frac{1}{f_{(\rm RN)}(r)}dr^{2}+r^{2}\left(
d\theta^{2}+\sin^{2}\theta d\phi^{2}
\right),\nonumber\\
\end{eqnarray}
\begin{eqnarray}
A_{({\rm RN})t}(r)&=&\frac{2Q}{r},
\end{eqnarray}
where
\begin{eqnarray}
f_{(\rm RN)}(r)&=&1-\frac{2M}{r}+\frac{Q^{2}}{r^{2}}.
\end{eqnarray}
Here, we assume $Q<M$.
In this spacetime, we obtain the expression
\begin{eqnarray}
L_{({\rm RN})}^{\mu\nu\alpha\beta}F_{({\rm RN})\mu\nu}F_{({\rm RN})\alpha\beta}&=&\frac{16Q^{2}(1-f_{(\rm RN)}(r))}{r^{6}}.
\end{eqnarray}
Since the BH event horizon is $r_{+}=M+\sqrt{M^{2}-Q^{2}}$, 
$L_{({\rm RN})}^{\mu\nu\alpha\beta}F_{({\rm RN})\mu\nu}F_{({\rm RN})\alpha\beta}$ is positive outside the horizon.
Therefore, if $H''(0)<0$, the scalar field must be constant under the RH BH.
\par
Let us see this statement from another point of view.
We consider a perturbation $\delta\Phi$ of the scalar field under the vanishing scalar profile.
For simplicity, we neglect the metric and vector perturbations.
From the Eq.\eqref{Eq:EOM scalar field}, the equation of motion for the perturbation is
\begin{eqnarray}
\nabla_{({\rm RN})\mu}\nabla_{({\rm RN})}^{\mu}\delta\Phi-m_{\rm eff}^{2}\delta\Phi=0,
\label{eq:EOM of the perturbation of scalar field}
\end{eqnarray}
where the effective mass $m_{\rm eff}$ is defined as
\begin{eqnarray}
m_{\rm eff}^{2}&:=&-H''(0)L_{(\rm RN)}^{\mu\nu\alpha\beta}F_{({\rm RN})\mu\nu}F_{({\rm RN})\alpha\beta}.
\end{eqnarray}
If $m_{\rm eff}^{2}$ is positive,
the scalar field is linearly stable, and it corresponds to the no-hair condition which is derived in the previous subsection.
On the other hand, if $m_{\rm eff}^{2}$ is negative, 
the scalar field suffers from a tachyonic instability, and it is expected that the solution with the non-trivial scalar field exists as the end point of the instability. 
\subsection{Stability analysis around the RN BHs}
We have investigated the test scalar field around the RN BH, and shown that if $H''(0)>0$, the scalar field may become unstable.
In this subsection, we discuss the linear stability of the scalar field around a RN BH by checking the behavior of the effective potential.
\par
\begin{figure}[htbp]
  \centering
  \includegraphics[width=0.45\textwidth]{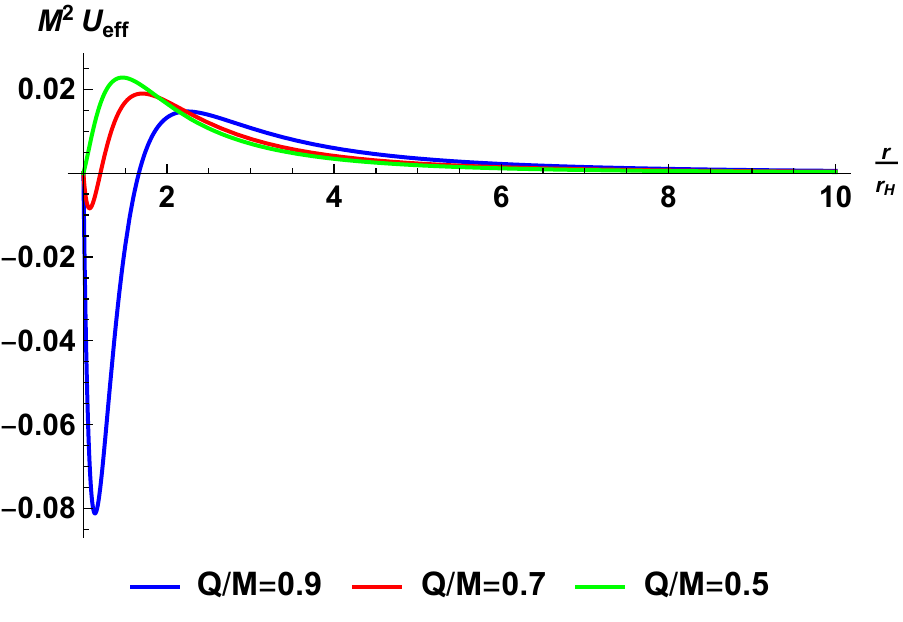}
  \caption{The effective potential for $H''(0)/M^{2}=2$ and $l=0$.}
  \label{Effective potential}
\end{figure}

Following the standard procedure, we separate the variables as follows:
\begin{eqnarray}
\Phi(t,r,\theta,\phi)&=&\sum_{l,m}\frac{\sigma_{lm}(t,r)}{r}Y_{lm}(\theta,\phi),
\end{eqnarray}
where $Y_{lm}(\theta,\phi)$ is the spherical harmonics.
From the Eq.\eqref{eq:EOM of the perturbation of scalar field}, we obtain the equation for $\sigma_{lm}$ as
\begin{equation}
-\frac{\partial^{2}\sigma_{lm}}{\partial t^{2}}+\frac{\partial^{2}\sigma_{lm}}{\partial r_{\ast}^{2}}=U_{\rm eff}\sigma_{lm},
\end{equation}
where $r_{\ast}$ is the tortoise coordinate of a RN BH:
\begin{eqnarray}
dr_{\ast}&=&\frac{dr}{f_{(\rm RN)}(r)},
\label{Eq:EOM for sigma}
\end{eqnarray}
and the effective potential $U_{\rm eff}$ is given as
\begin{eqnarray}
U_{\rm eff}=f_{(\rm RN)}(r)&&\left(
-\frac{16H''(0)Q^{2}(1-f_{(\rm RN)}(r))}{r^{6}}
\right.\nonumber\\
&&\left.
+\frac{l(l+1)}{r^{2}}
+\frac{f_{(\rm RN)}'(r)}{r}
\right),\nonumber\\
\end{eqnarray}
where $f'_{(\rm RN)}(r)=\frac{d}{dr}f_{(\rm RN)}(r)$.
The typical profile of the effective potential is depicted in Fig.\ref{Effective potential}.
As expected, there is the region in which the effective potential is negative around the event horizon, 
and it implies that there may be an unstable scalar mode.
\par
\begin{figure}[htbp]
  \centering
  \includegraphics[width=0.45\textwidth]{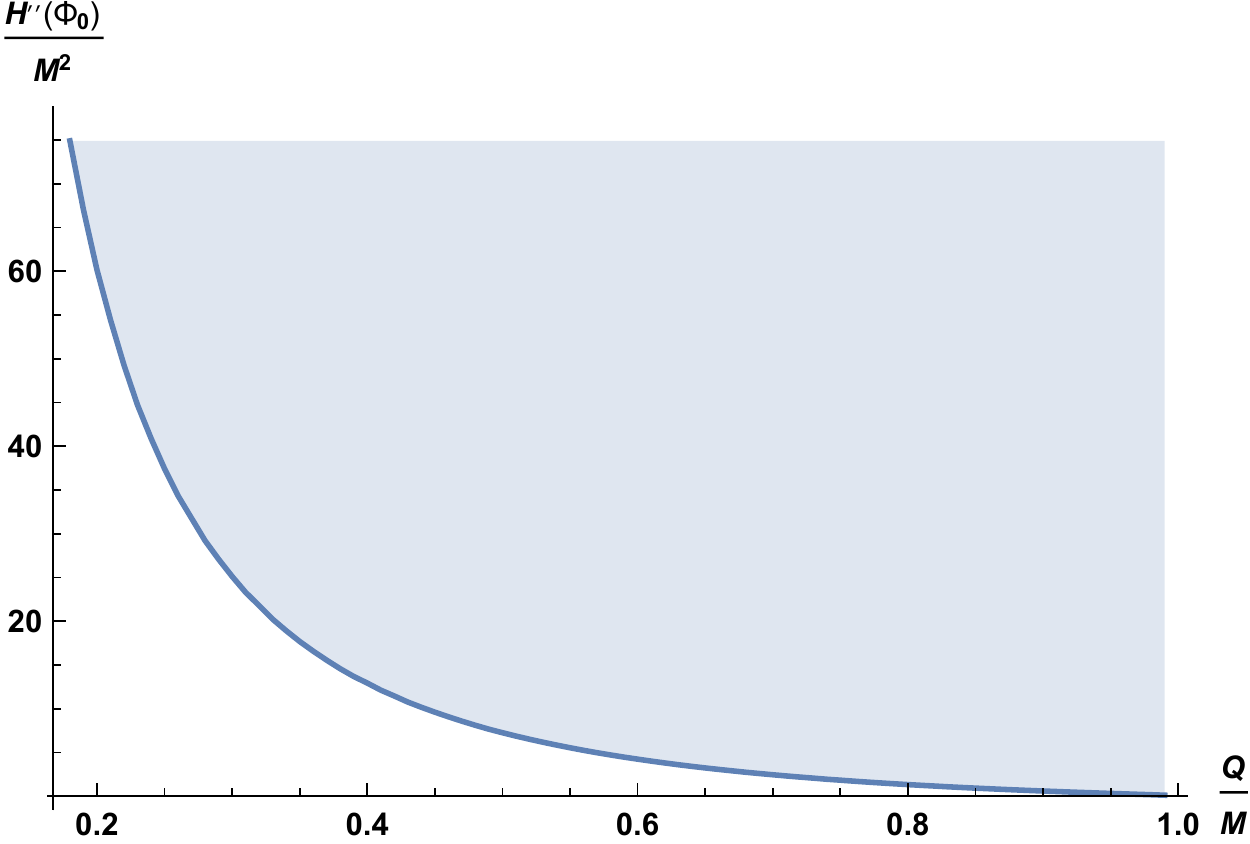}
  \caption{The parameter region in which the constant scalar field solution on RN BHs is unstable for the $l=0$ mode of perturbations.
  The solutions in the shaded region are unstable.}\label{graph:RN_instability_bdry_Sdeformation}
\end{figure}

To verify the existence of RN BH solutions for which the stability is not ensured,
we use the method which is discussed in\cite{Kimura:2017uor}. 
In the frequency domain, the equation of motion Eq.\eqref{Eq:EOM for sigma} becomes
\begin{eqnarray}
-\frac{d^{2}}{dr_{\ast}^{2}}\tilde\sigma_{lm\omega}+U_{\rm eff}(r)\tilde{\sigma}_{lm\omega}=\omega^{2}\tilde{\sigma}_{lm\omega},
\label{Eq:EOM for sigma in frequency domain}
\end{eqnarray}
where $\sigma_{lm}(t,  r)=e^{i\omega t}\tilde{\sigma}_{lm\omega}(r)$.
To investigate the linear stability which means that $\omega^{2}$ is nonnegative for all modes, 
we focus on the following equation for $\varphi(r_{\ast})$:
\begin{eqnarray}
\left(
-\frac{d^{2}}{dr_{\ast}^{2}}+U_{\rm eff}
\right)\varphi=0.\label{eq:Kimura eq}
\end{eqnarray}
As discussed in \cite{Kimura:2017uor}, 
if the solution $\varphi(r_{\ast})$ of Eq.\eqref{eq:Kimura eq} does not cross zero, 
the scalar field is linearly stable.
Using the method, we found that there is a parameter region in which the constant scalar field solution on a RN BH can be unstable (see Fig.\ref{graph:RN_instability_bdry_Sdeformation}).
\section{Test field analysis}\label{Sec. Test field analysis}
So far, we have investigated the stability of the constant scalar profile around the RN BH,
and found that there are parameter regions in which the $l=0$ mode of the scalar field perturbation is unstable.
This result implies that there are bound states for a negative effective potential in the unstable parameter region.
We expect that the state is the final state of the instability.
In this section, we construct the profile of the bound state and investigate its parameter dependence.
\footnote{
The test field analysis in spontaneous scalarization for the Einstein-scalar-Gauss-Bonnet theory was used in \cite{Minamitsuji:2018xde}, and it was shown that the stability analysis of the test field is qualitatively same as the stability of the hairy BH.
Furthermore, the validity of the test field analysis and relation between the analysis and the hairy BH is discussed in Appendix\ref{appendix:Validity of the test field analysis}.
}
\subsection{Bound states of the test scalar field around the RN BHs.}
The static and spherically symmetric profiles of the scalar field around a RN BH are solutions of the following equation:
\begin{widetext}
\begin{eqnarray}
\partial_{r}^{2}\Phi
+\left(
\frac{2}{r}
+\frac{f'_{(\rm RN)}(r)}{f_{(\rm RN)}(r)}
\right)\partial_{r}\Phi
+16Q^{2}\frac{H'(\Phi(r))}{r^{6}}\left(
-1+\frac{1}{f_{(\rm RN)}(r)}
\right)=0.
\end{eqnarray}
\end{widetext}
We can solve this equation with the regularity boundary condition on the event horizon.
Assuming that the bound state is the final state of the instability around the vanishing profile $\Phi=0$, 
we impose $\Phi(r\to \infty)=0$ as the boundary condition.
The scalar charge $Q_{\rm s}$ is defined as
\begin{eqnarray}
Q_{\rm s}=-r^{2}\Phi'(r)|_{r\to\infty}.
\end{eqnarray}
\par
From here, we specify the coupling function $H(\Phi)$.
As a first step, we consider the simplest coupling:
\begin{eqnarray}
H(\Phi)&=&\frac{\eta}{2}\Phi^{2},
\end{eqnarray}
where $\eta$ is a constant.
This model is named the quadratic model.
As we will show in the next subsection, the bound state of this coupling is unstable.
In order to stabilize the bound state, the nonlinear effect of $H(\Phi)$ plays an important role.
In particular, the existence of the region in which $H''(\Phi)$ is negative is crucial.
Then, we also consider the following coupling function as an example model:
\begin{eqnarray}
H(\Phi)=\frac{\eta}{2}\left(1-e^{-\Phi^{2}}\right).
\end{eqnarray}
This model is named the exponential model.
\subsubsection{The quadratic model}
\begin{figure}[htbp]
  \centering
  \includegraphics[width=0.45\textwidth]{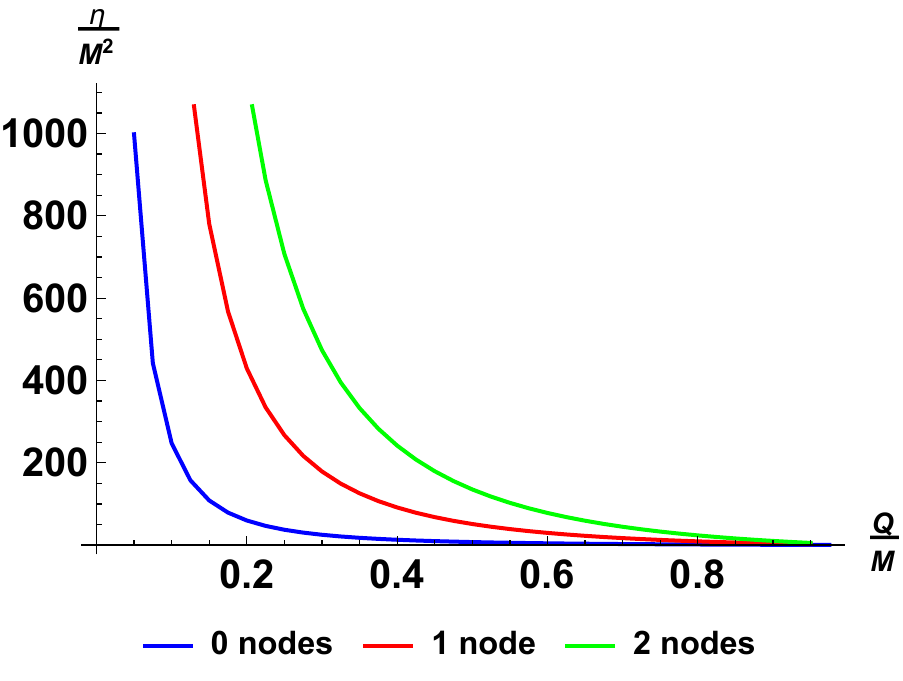}
  \caption{The relation between the coupling $\eta/M^{2}$ and the electric charge $Q/M$ for different number of nodes in the quadratic model.}
  \label{graph_Q_eta_nontrivial_scalar_on_RN}
\end{figure}

In the quadratic model, the equation of motion for the scalar field becomes linear, and the normalization factor of the field is irrelevant as long as the backreaction can be ignored.
We found that for each electric charge $Q$, there is a solution which is characterized by the number of nodes.
\begin{figure}[htbp]
  \centering
  \includegraphics[width=0.45\textwidth]{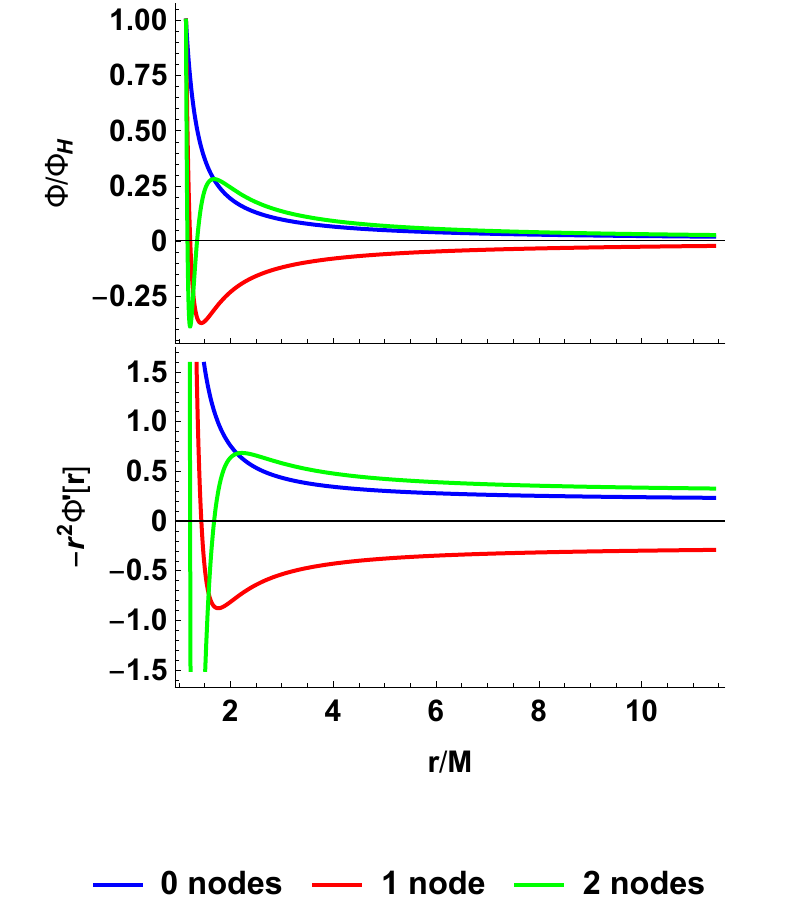}
  \caption{The typical behavior of the scalar field for the quadratic model. The electric charge is fixed as $Q/M=0.99$.
  The corresponding parameter sets are summarized in Table.\ref{The typical parameter set of the bound state in quadratic model}.
  }\label{graph_typical_behavior_nontrivial_scalar_on_RN}
\end{figure}
\begin{table}
\begin{tabular}{|c||c|c|}
\hline
&$\eta/M^{2}$&$Q_{s}/(M\Phi_{\rm H})$\\
\hline\hline
0 nodes&0.128&0.193\\
\hline
1 node&0.661&-0.243\\
\hline
2 nodes&1.669&0.271\\
\hline
\end{tabular}
\caption{The typical parameter set of the bound states in the quadratic model with $Q/M=0.99$.}
\label{The typical parameter set of the bound state in quadratic model}
\end{table}
The typical profiles of the scalar field are depicted  in Fig.\ref{graph_typical_behavior_nontrivial_scalar_on_RN}, and the corresponding parameter sets are summarized in Table.\ref{The typical parameter set of the bound state in quadratic model}.
In the figure, $\Phi$ is normalized such that the scalar field on the horizon is unity.
Fig.\ref{graph_Q_eta_nontrivial_scalar_on_RN} shows the relation between the electric charge $Q/M$ and the coupling $\eta/M^{2}$ for each number of nodes.
From the plot, we find that there are solutions with 0,1 and 2 nodes for any $Q/M$, and there may be solutions with any nodes for any $Q/M$.
We found that the line of the solution with 0 nodes in the figure is almost the same as the critical line of Fig.\ref{graph:RN_instability_bdry_Sdeformation}.
\subsubsection{The exponential model}
\begin{figure}[htbp]
      \includegraphics[width=0.45\textwidth]{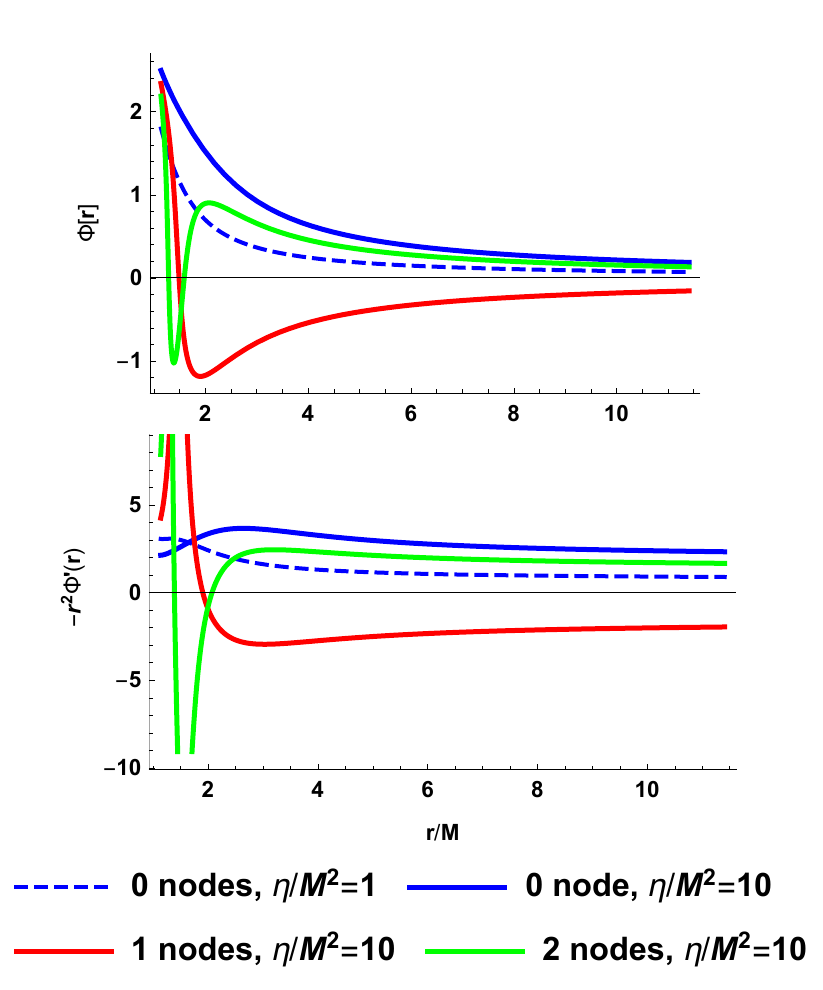}\\
   \caption{The typical behavior of the non-trivial profile of the scalar field around a RN BH for $Q/M=0.99$ in the exponential model.
  The corresponding parameter sets are summarized in Table\ref{The typical parameter set of the bound state in exponential model}.}
  \label{graph_typical_profile_non-trivial_profile_RN}
\end{figure}

\begin{figure}[htbp]
  \centering
  \includegraphics[width=0.45\textwidth]{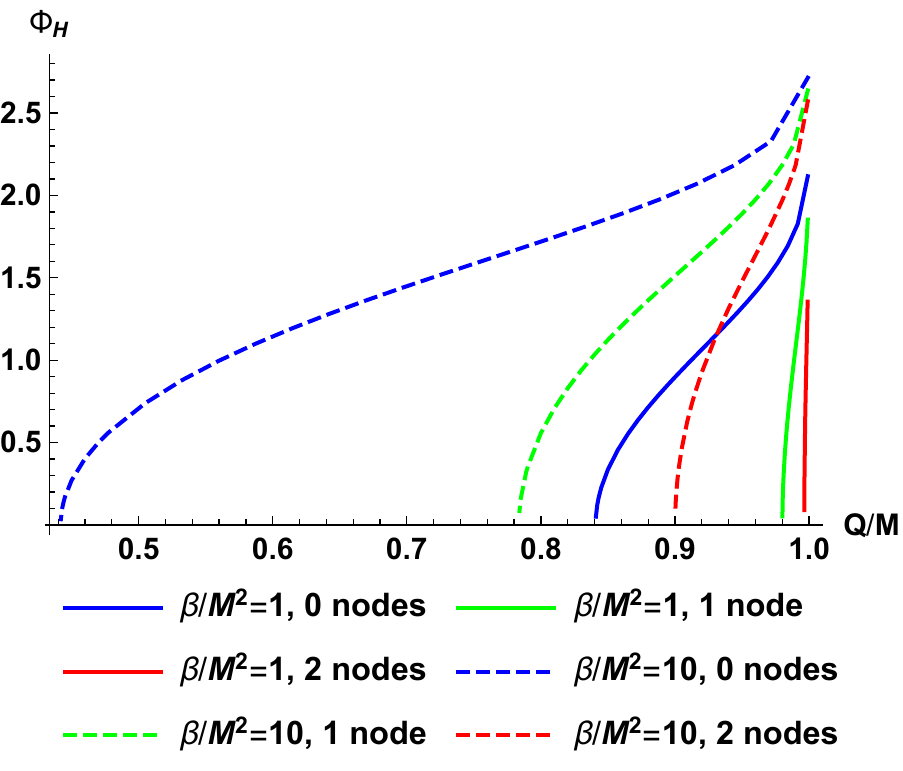}
  \caption{The relation between $Q/M$ and $\Phi_{\rm H}$ for the nontrivial profile of the scalar field on RN BHs for the exponential model.
  The graph shows that there is a critical value of $Q/M$ below which the solution does not exist for a fixed value of $\eta/M^{2}$.}\label{graph_Q_Phih_non-trivial_profile_RN}
\end{figure}

In the exponential model, the equation of motion for the scalar field is nonlinear, and the relevant free parameters are $\eta/M^{2}$,$Q/M$,$\Phi_{H}$, where $\Phi_{H}$ is the value of the scalar field on the event horizon.
The typical profile of the scalar field is depicted in Fig.\ref{graph_typical_profile_non-trivial_profile_RN}, and the corresponding parameter sets are summarized in Table.\ref{The typical parameter set of the bound state in exponential model}
\begin{table}
\begin{tabular}{|c||c|c|}
\hline
&$\Phi_{\rm H}$&$Q_{\rm s}/M$\\
\hline\hline
0 nodes, $\eta/M^{2}=1$&$1.80$&$0.734$\\
\hline
0 nodes, $\eta/M^{2}=10$&$2.49$&$1.93$\\
\hline
1 node, $\eta/M^{2}=10$&$2.34$&$-1.64$\\
\hline
2 nodes, $\eta/M^{2}=10$&$2.19$&$1.38$\\
\hline
\end{tabular}
\caption{The typical parameter set of the bound states in the exponential model with $Q/M=0.99$.}
\label{The typical parameter set of the bound state in exponential model}
\end{table}
For fixed coupling $\eta/M^{2}$ and electric charge, $\Phi_{\rm H}$ becomes large as the number of node decreases.
As $\eta/M^{2}$ or $Q/M$ is increased, the solution with higher nodes can be constructed.
Figure \ref{graph_Q_Phih_non-trivial_profile_RN} shows the relation between $\Phi_{H}$ and the charge-to-mass ratio $Q/M$ for different number of nodes and coupling $\eta/M^{2}$.
From the plot, for fixed coupling $\eta/M^{2}$ and nodes, $\Phi_{H}$ is an increasing function for the charge-to-mass ratio.
There is the critical value $Q_{\rm c}(\eta/M^{2},n)/M$ for a solution with $n$ nodes in which $\Phi_{H}$ vanishes.
It means that if $Q/M<Q_{\rm c}(\eta/M^{2},n)/M$, the solution with $n$ nodes does not exist.
We have scanned the parameter region of $(Q/M,\eta/M^{2})$ in which the bound state exists, and found that the critical line above which there is the bound state almost coincides with the line of Fig.\ref{graph:RN_instability_bdry_Sdeformation}.
In this model, the scalar charge can be of order one, which means that the large difference from GR appears in the asymptotic region.
\subsection{Stability analysis of the bound state}
\label{stability_bound_state}
We discuss the stability of the bound state, which is constructed in the previous subsection, around a RN BH.
Let us denote the bound state as $\Phi_{0}(r)$, and consider the perturbation around the state:
\begin{eqnarray}
\Phi(t,r,\theta,\phi)&=&\Phi_{0}(r)+\sum_{l,m}\frac{\sigma_{lm}(t,r)}{r}Y_{lm}(\theta,\phi),
\end{eqnarray}
where $Y_{lm}(\theta,\phi)$ is the spherical harmonics.
From the equation of motion for the scalar field, we obtain the equation for $\sigma_{lm}$ as
\begin{equation}
-\frac{\partial^{2}\sigma_{lm}}{\partial t^{2}}+\frac{\partial^{2}\sigma_{lm}}{\partial r_{\ast}^{2}}=U_{\rm eff}\sigma_{lm},
\end{equation}
where the effective potential $U_{\rm eff}$ is given as
\begin{eqnarray}
U_{\rm eff}=f_{(\rm RN)}(r)&&\left(
-\frac{16H''(\Phi_{0}(r))Q^{2}(1-f_{(\rm RN)}(r))}{r^{6}}
\right.\nonumber\\
&&\left.
+\frac{l(l+1)}{r^{2}}
+\frac{f_{(\rm RN)}'(r)}{r}
\right).\label{effective_potential}
\end{eqnarray}
\subsubsection{The quadratic model}
In the quadratic model, the effective potential is the same as one of the perturbations around the constant scalar field, and hence the stability condition remains the same.
Therefore, the bound state with $H(\Phi)=\frac{\eta}{2}\Phi^{2}$ is unstable against the radial perturbations.
\subsubsection{The exponential model}
\begin{figure}[htbp]
  \centering
  \includegraphics[width=0.45\textwidth]{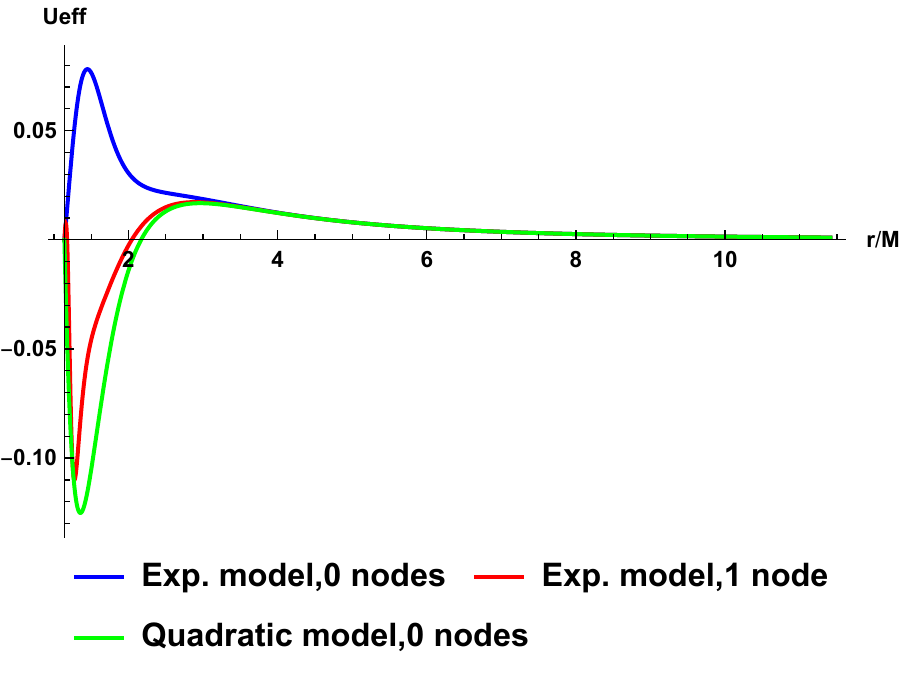}
  \caption{The effective potential for $Q/M=0.99,\eta/M^{2}=1$.
  The blue line corresponds to nodeless solution for the exponential mode.
  The red line corresponds to the solution with 1 node.
  The green line corresponds to nodeless solution for the quadratic model.
  }\label{graph_Ueff_non-trivial_profile_RN_BH}
\end{figure}

In the exponential model, the effective potential of the radial perturbation around the bound state on the RN BH is shown in Fig.\ref{graph_Ueff_non-trivial_profile_RN_BH}.
The positive effective potential ensures the stability of the bound state.
From Eq.\eqref{effective_potential}, in order that the effective potential becomes positive everywhere, 
$H''(\Phi_{0}(r))$ must be negative around the horizon.
The condition in which $H''(\Phi_{0})$ is negative around the horizon is $\frac{1}{\sqrt{2}}<|\Phi_{0}|$.
Figure \ref{graph_Ueff_non-trivial_profile_RN_BH} implies that there is parameter sets in which the effective potential becomes positive everywhere.
Therefore, the nonlinearity of $H(\Phi)$ can stabilize the bound state.
\section{Scalarized solution}
In this section, we construct the scalarized BH solutions as the final state of spontaneous scalarization and discuss the stability.
\subsection{Construction}
To construct the solutions, let us focus on the equations of motion.
Due to the $U(1)$ symmetry, the equation of motion for the vector field can be expressed in the divergence form:
\begin{eqnarray}
\nabla_{\mu}\left(
F^{\mu\nu}-4H(\Phi)L^{\alpha\beta\mu\nu}F_{\alpha\beta}
\right)=0.
\label{eq:Maxwell equation}
\end{eqnarray}
We assume that the spacetime is spherically symmetric, whose metric is given as
\begin{eqnarray}
ds^{2}=-a(r)dt^{2}+\frac{dr^{2}}{b(r)}+r^{2}(d\theta^{2}+\sin^{2}\theta d\phi^{2}),
\end{eqnarray}
and $\Phi=\Phi(r)$.
Varying the action with respect to the metric and the scalar field, we have the set of the ordinary differential equations:
\begin{widetext}
\begin{eqnarray}
b'-\frac{1-b}{r}+\frac{rb\Phi'^{2}}{4}+\frac{rbA_{t}'^{2}}{4a}+\frac{2bH(\Phi)A_{t}'^{2}}{ra}\left(1-b\right)=0,
\label{eq:equation for B}\\
\frac{a'}{a}-\frac{1-b}{rb}-\frac{r\Phi'^{2}}{4}+\frac{rA_{t}'^{2}}{4a}-\frac{2H(\Phi)A_{t}'^{2}}{ra}(-1+3b)=0,
\label{eq:equation for A}\\
\Phi''+\left(
\frac{2}{r}+\frac{a'}{2a}+\frac{b'}{2b}
\right)\Phi'
+\frac{4A_{t}'^{2}H'(\Phi)}{r^{2}a}(1-b)=0.
\label{eq:equation for phi}
\end{eqnarray}
\end{widetext}
Since the equation for the vector field is written as the divergence form as Eq.\eqref{eq:Maxwell equation},
we can integrate it and obtain
\begin{eqnarray}
A_{t}'=-\sqrt{\frac{a}{b}}\frac{2Q}{r^{2}-8(b-1)H(\Phi)},
\label{eq: equation for At}
\end{eqnarray}
where $Q$ is an integration constant.
\par
We substitute Eq.\eqref{eq: equation for At} for Eqs.\eqref{eq:equation for B}-\eqref{eq:equation for phi}.
Then, we have three equations for $a$, $b$, and $\Phi$, and can solve them numerically.
In the vicinity of the horizon, the metric functions behave as $a(r),b(r)\propto (r-r_{\rm H})$.
From the regularity condition, the behavior of the fields around the horizon is given by:
\begin{widetext}
\begin{eqnarray}
\Phi(r)&=&\Phi_{\rm H}-\frac{16Q^{2}H'(\Phi_{\rm H})}{r_{\rm H}(r_{\rm H}^{2}+8H(\Phi_{\rm H}))(-Q^{2}+r_{\rm H}^{2}+8H(\Phi_{\rm H}))}(r-r_{\rm H})+\mathcal{O}\left((r-r_{\rm H})^{2}\right),
\label{eq:phi around the horizon}\\
a(r)&=&a_{1}(r-r_{\rm H})+\mathcal{O}\left((r-r_{\rm H})^{2}\right),
\label{eq:A around the horizon}\\
b(r)&=&\frac{-Q^{2}+r_{\rm H}^{2}+8H(\Phi_{\rm H})}{r_{\rm H}(r_{\rm H}^{2}+8H(\Phi_{\rm H}))}(r-r_{\rm H})+\mathcal{O}\left((r-r_{\rm H})^{2}\right),
\label{eq:B around the horizon}
\end{eqnarray}
\end{widetext}
where $\Phi_{\rm H}$ is the value of the scalar field on the event horizon, and $a_{1}$ is a free parameter.
\par
We impose the asymptotically flat boundary condition for the metric, 
and the vanishing vector and scalar fields at the infinity.
From the equations, the asymptotic behaviors of the fields are given as
\begin{eqnarray}
\Phi(r)&=&\frac{Q_{\rm s}}{r}+\frac{MQ_{\rm s}}{r^{2}}+\mathcal{O}\left(\frac{1}{r^{3}}\right),\\
a(r)&=&1-\frac{2M}{r}+\frac{Q^{2}}{r^{2}}+\mathcal{O}\left(\frac{1}{r^{3}}\right),\\
b(r)&=&1-\frac{2M}{r}+\frac{4Q^{2}+Q_{\rm s}^{2}}{4r^{2}}+\mathcal{O}\left(\frac{1}{r^{3}}\right),\\
A_{t}(r)&=&\frac{2Q}{r}+\mathcal{O}\left(\frac{1}{r^{3}}\right),\label{At at infinity}
\end{eqnarray}
where $M$ and $Q_{\rm s}$ are integration constants, which correspond to mass and scalar charge.
From Eq.\eqref{At at infinity}, $Q$ corresponds to the electric charge.
\par
We numerically integrate the equations of motion under the boundary conditions around the horizon and at the infinity,
and construct the scalarized BH solutions in the quadratic and the exponential models.
\subsubsection{The quadratic model}
\begin{figure}[htbp]
  \centering
  \includegraphics[width=0.45\textwidth]{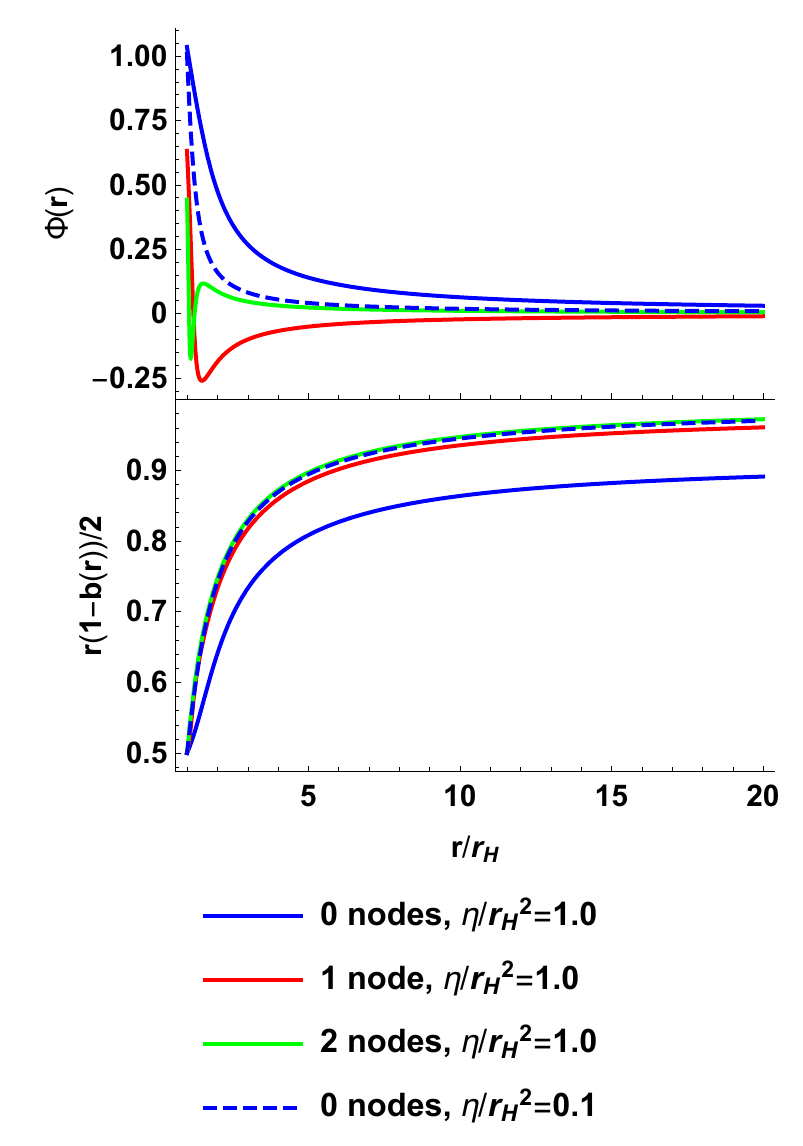}
  \caption{The typical profile of the scalarized BH solutions for $Q/r_{\rm H}=1.0$ for quadratic model.
  The corresponding parameter sets of each solutions are summarized in Table\ref{table:scalarized BH quadratic model}.
  }\label{graph_typical_behavior_hairyBH}
\end{figure}

The typical profile of the fields is depicted in Fig.\ref{graph_typical_behavior_hairyBH}, and 
the parameters of the solutions are summarized in Table.\ref{table:scalarized BH quadratic model}.
\begin{figure}[htbp]
  \centering
  \includegraphics[width=0.45\textwidth]{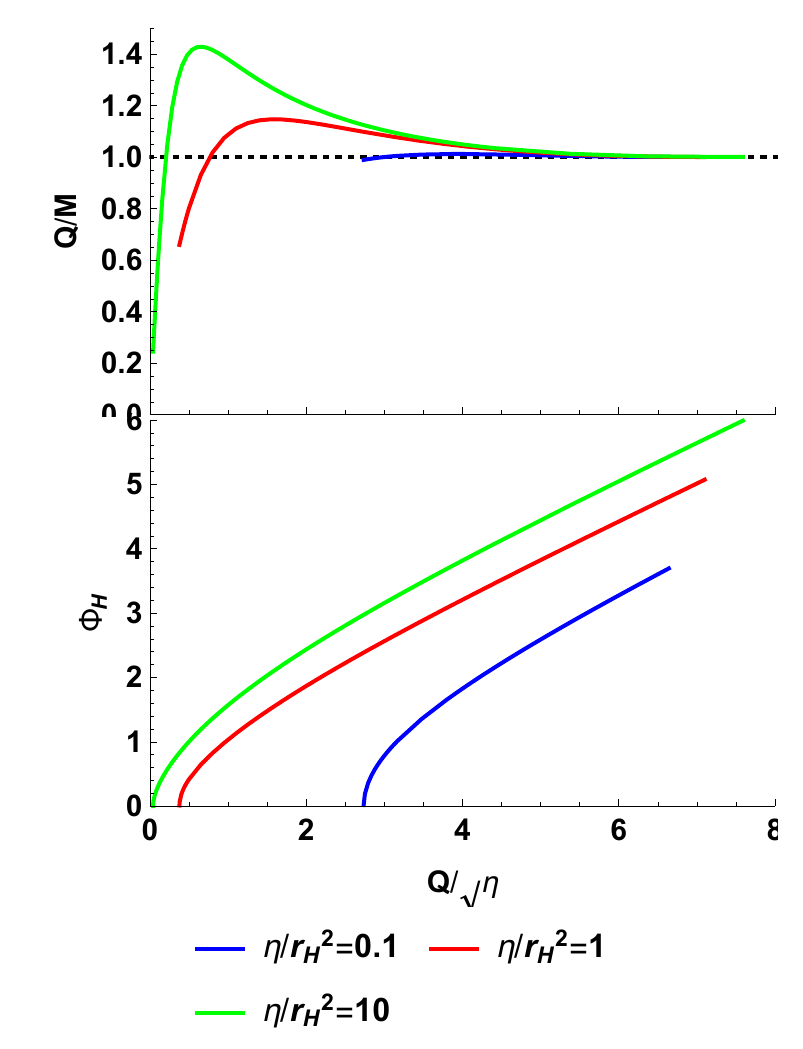}
  \caption{The sequence of the scalarized BH solutions for fixed value $\eta/r_{\rm H}^{2}$ in the quadratic model.
  Top : The relation between $Q/\sqrt{\eta}$ and $Q/M$ for nodeless solution.
  Dotted line is $Q/M=1$.
  Bottom : The relation between $Q/\sqrt{\eta}$ and the scalar field on the horizon.
  }\label{graph_hairyBH_relation_Q_QM}
\end{figure}
\begin{table}
\begin{tabular}{|c||c|c|c|}
\hline
&$Q/M$&$Q_{s}/M$\\
\hline
\hline
0 nodes,$\eta/r_{\rm H}^{2}=1.0$&$1.09$&$0.622$\\
\hline
1 node,$\eta/r_{\rm H}^{2}=1.0$&$1.01$&$-0.209$\\
\hline
2 nodes,$\eta/r_{\rm H}^{2}=1.0$&$1.00$&$0.0929$\\
\hline
0 nodes,$\eta/r_{\rm H}^{2}=0.1$&$1.01$&$0.164$\\
\hline
\end{tabular}
\caption{The typical parameter sets for the scalarized BH solutions for the quadratic model.}
\label{table:scalarized BH quadratic model}
\end{table}
These solutions can be overcharged.
We found that the 0 nodes solution has the biggest $Q/M$ for fixed $\eta/M^{2}$ and $Q/\sqrt{\eta}$.
Since nodeless solutions are more likely to be stable, we focus on nodeless solutions.
The relation between $Q/M$ and $Q/\sqrt{\eta}$ and the relation between $\Phi_{\rm H}$ and $Q/\sqrt{\eta}$ in nodeless solution are plotted in Fig.\ref{graph_hairyBH_relation_Q_QM}.
The figure shows that $Q/M$ as the function of $Q/\sqrt{\eta}$ has a maximum value, and $Q/M$ approaches $1$ for larger $Q/r_{H}$.
Furthermore, for the large coupling $\eta/r_{H}^{2}$, the maximum value of the charge-to-mass ratio becomes large.
There is a critical parameter $Q_{c}/\sqrt{\eta}$ for $Q/\sqrt{\eta}$, above which we can not construct the solution, numerically.
We could not determine whether there are solutions above the critical parameter, or not.
On the other hand, there is also the critical value below which the scalarized BH solution does not exist, and on which the scalar field on the event horizon becomes zero.
\par
In the case of the scalarization induced by the term $F^{\mu\nu}F_{\mu\nu}$, there is the critical line in the parameter region of the coupling and $Q/M$\cite{Herdeiro:2018wub,Fernandes:2019rez}.
For a fixed value of coupling, there is no scalarized BH solution whose charge-to-mass ratio is larger than the critical line, on which the solution is extremal.
On the other hand, in our model, the BH solution which has the maximum $Q/M$ is not the extremal BH for a fixed value of $\eta/r_{\rm H}^{2}$.
It is expected that there is the critical line on the parameter plane of $(\frac{\eta}{r_{\rm H}^{2}},\frac{Q}{M})$, which corresponds to the maximum value of $Q/M$ for a fixed value of coupling.
We leave it as a future work.
\par
Finally, we shortly mention the expected properties of the extremal BH in the $f_{4}$ model.
From the Eqs.\eqref{eq:phi around the horizon}-\eqref{eq:B around the horizon}, the extremal BH, in which the coefficient of $r-r_{\rm H}$ vanishes, must satisfy the following condition:
\begin{eqnarray}
-Q^{2}+r_{\rm H}^{2}+8H(\Phi_{\rm H})=0.
\end{eqnarray}
When this condition is satisfied, the denominator of the coefficient of $r-r_{\rm H}$ in $\Phi$ vanishes.
To construct the extremal BH solution with finite scalar field, there are two possibilities.
One possibility is that $H'(\Phi_{\rm H})$ is zero, and one particular solution is $\Phi_{\rm H}=0$.
Then, the solution does not have a nontrivial scalar field.
The other possibility is that our ansatz of expansion for the field (i.e.Eqs.\eqref{eq:phi around the horizon}-\eqref{eq:B around the horizon}) is not valid.
It would be possible to construct the extremal BH by generalization of the ansatz.
We also leave the construction of the extremal BH in the $f_{4}$ model as a future work.
\subsubsection{The exponential model}
\begin{figure}[htbp]
  \centering
  \includegraphics[width=0.45\textwidth]{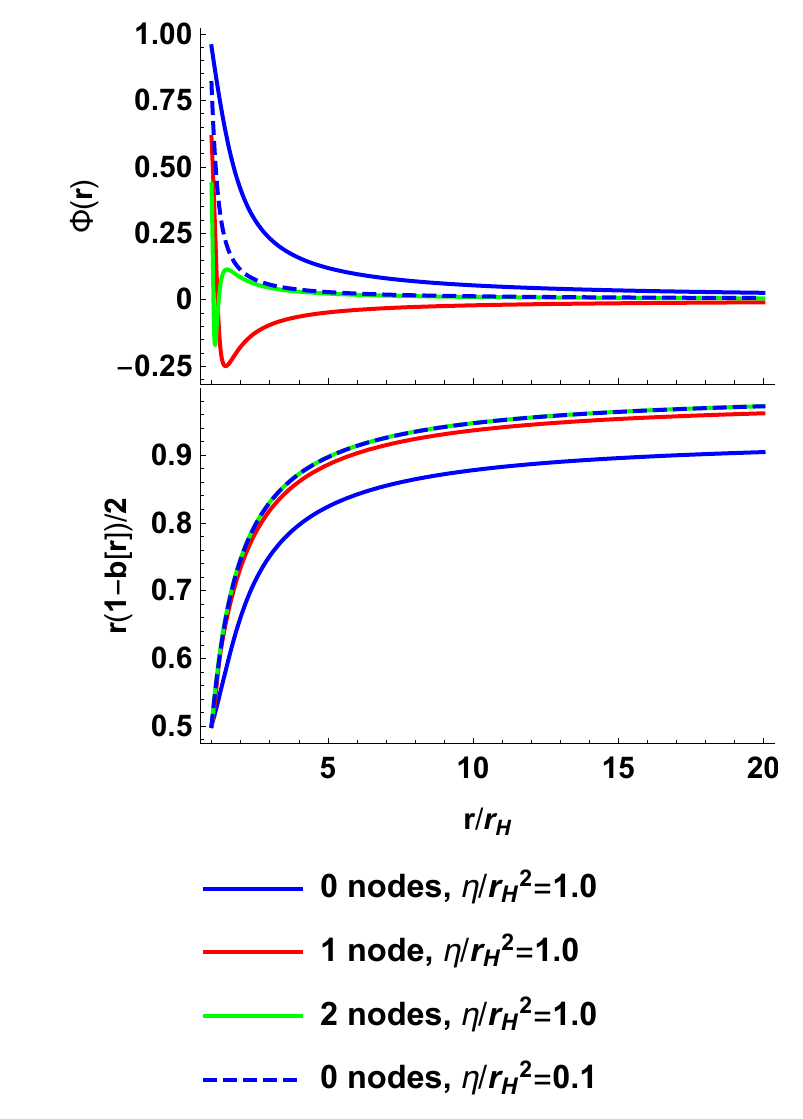}
  \caption{The typical behavior of the scalarized BH solutions with $Q/r_{\rm H}=1$ in the exponential model.
   The corresponding parameter sets are summarized in Table\ref{table:scalarized BH for exponential model}.
  }\label{graph_typical_behavior_hairyBH_exp}
  \end{figure}

The typical behavior of the fields is depicted in Fig.\ref{graph_typical_behavior_hairyBH_exp}.
\begin{figure}[htbp]
  \centering
  \includegraphics[width=0.45\textwidth]{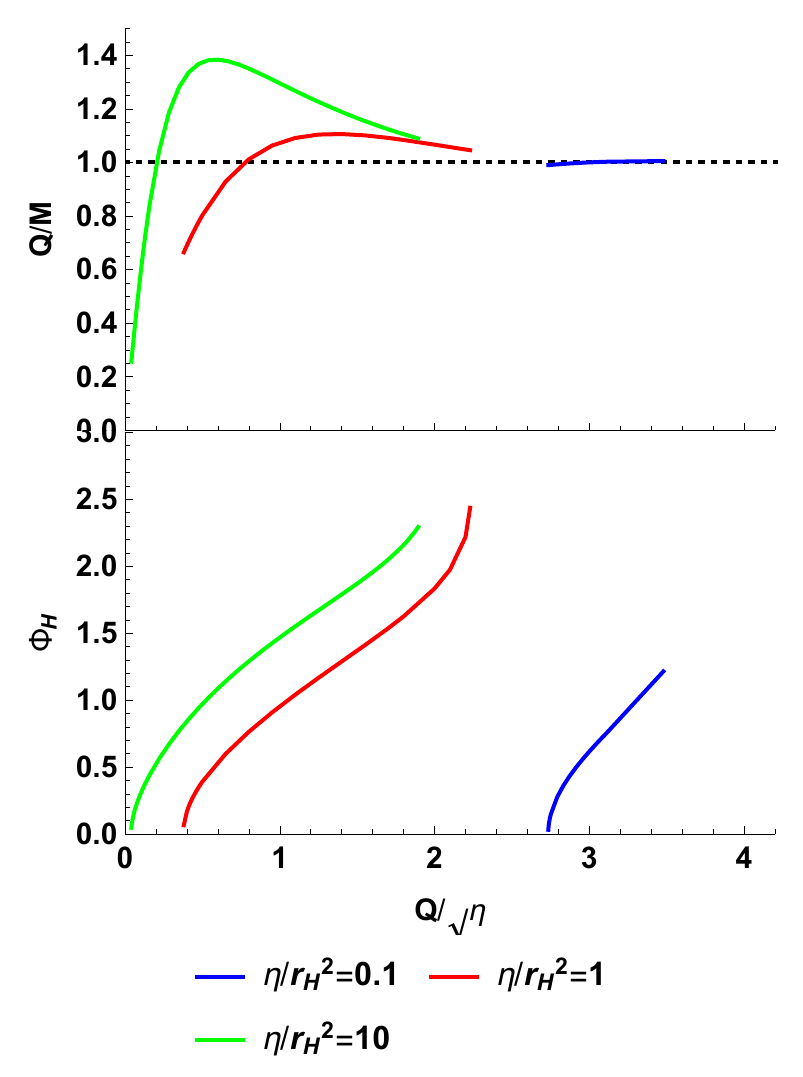}
  \caption{The sequence of the scalarized BH solutions for the fixed value $\eta/r_{\rm H}^{2}$ in the exponential model.
  Top : The relation between $Q/\sqrt{\eta}$ and $Q/M$ for node-less solution.
  Dotted line is $Q/M=1$.
  Bottom : The relation between $Q/\sqrt{\eta}$ and the scalar field on the horizon.
  }\label{graph_hairyBH_relation_Q_QM_exp}
\end{figure}
\begin{table}
\begin{tabular}{|c||c|c|c|}
\hline
&$Q/M$&$Q_{s}/M$\\
\hline
\hline
0 nodes,$\eta/r_{\rm H}^{2}=1.0$&$1.07$&$0.523$\\
\hline
1 nodes,$\eta/r_{\rm H}^{2}=1.0$&$1.01$&$-0.197$\\
\hline
2 nodes,$\eta/r_{\rm H}^{2}=1.0$&$1.00$&$0.0902$\\
\hline
0 nodes,$\eta/r_{\rm H}^{2}=0.1$&$1.00$&$0.113$\\
\hline
\end{tabular}
\caption{The typical parameter sets for the scalarized BH solutions for the exponential model.}
\label{table:scalarized BH for exponential model}
\end{table}
These behaviors are almost the same as the quadratic model except around the horizon.
The relation between $Q/M$ and $Q/\sqrt{\eta}$ is depicted in Fig.\ref{graph_hairyBH_relation_Q_QM_exp}.
The critical parameter above which we could not construct the solutions numerically is smaller than the quadratic case,
and we could not determine whether solutions exit beyond the critical value, again.
Furthermore, there is a solution whose $\Phi_{\rm H}$ becomes zero for each value of $\eta$.
If the $Q/\sqrt{\eta}$ is smaller than the value of the solution, there is no scalarized BH solution.
\subsection{Stability}
\begin{figure}[htbp]
  \centering
  \includegraphics[width=0.45\textwidth]{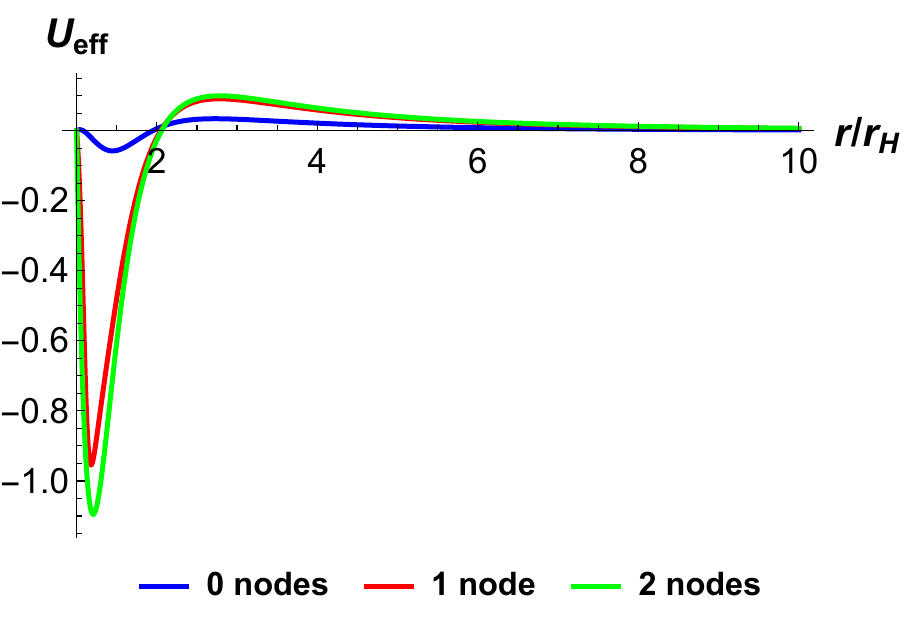}
  \caption{The effective potential for the scalarized BH solutions $l=0, Q/\sqrt{\eta}=0.9, \eta/r_H^2=1$ in the quadratic model.
  }\label{graph_effective_potential_hairy_quad}
\end{figure}

\begin{figure}[htbp]
  \centering
  \includegraphics[width=0.45\textwidth]{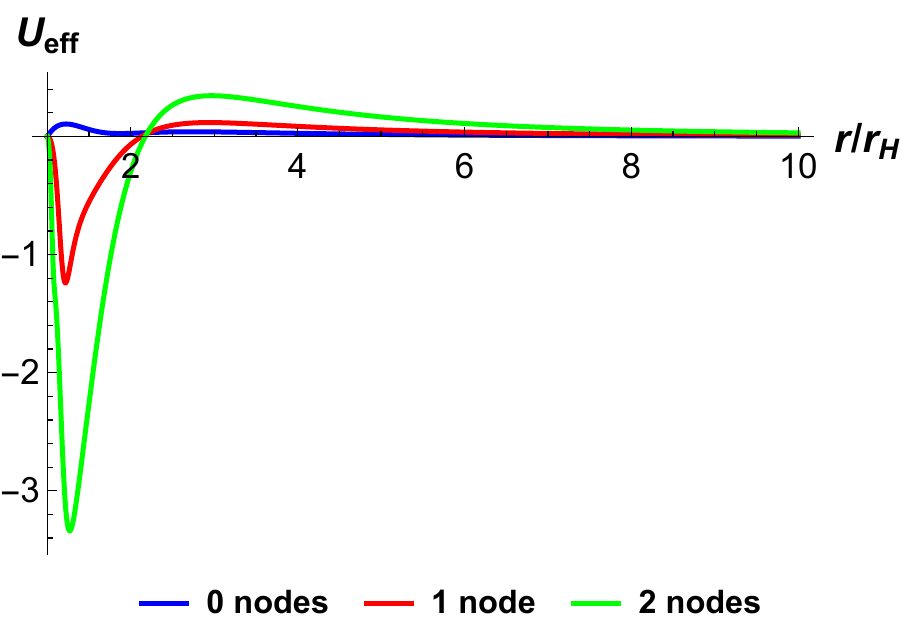}
  \caption{The effective potential for the scalarized BH solutions $l=0, Q/\sqrt{\eta}=1.0, \eta/r_H^2=1$ in the exponential model.
  }\label{graph_effective_potential_hairy_exp}
\end{figure}

We can check the stability of the scalarized BH solutions by calculating the effective potential.
Let us consider the perturbation around the scalarized BH solutions $\Phi_{\rm s}$, which is constructed by solving Eqs.\eqref{eq:equation for B}-\eqref{eq:equation for phi}.
Here, we only consider the stability against the perturbation of the scalar field:
\begin{eqnarray}
\Phi(t,r,\theta,\phi)&=&\Phi_{\rm s}(r)+\sum_{l,m}\frac{\sigma_{lm}(t,r)}{r}Y_{lm}(\theta,\phi).
\end{eqnarray}
From the equation of motion for the scalar field, we have the equation for $\sigma_{lm}$:
\begin{equation}
-\frac{\partial^{2}\sigma_{lm}}{\partial t^{2}}+\frac{\partial^{2}\sigma_{lm}}{\partial r_{\ast}^{2}}=U_{\rm eff}\sigma_{lm},
\end{equation}
where $dr_{\ast}=dr/\sqrt{ab}$.
The effective potential $U_{\rm eff}(r)$ is given as
\begin{eqnarray}
U_{\rm eff}&&=\frac{l(l+1)}{r^2}a+\frac{1}{2r}(a'b+ab')\nonumber\\
&&~~-H''(\Phi_{\rm s})\frac{16Q^2a(1-b)}{r^2(r^2+8(1-b)H(\Phi_{\rm s}))^2}.
\end{eqnarray} 
The typical form of the effective potential for each model is plotted in Fig. \ref{graph_effective_potential_hairy_quad} and Fig. \ref{graph_effective_potential_hairy_exp}.
In the exponential model, the effective potential for the nodeless solution is positive everywhere outside the horizon as shown in Fig.\ref{graph_effective_potential_hairy_exp}.
As is expected from the bound state analysis in Sec. \ref{stability_bound_state}, the solution is stable for the $l=0$ mode scalar perturbation.
Therefore, we conclude the scalarized BH solution in the exponential model can be the end state of scalarization. 
\section{Summary and discussion}
In this paper, we have investigated spontaneous scalarization of charged BHs in the scalar-vector-tensor(SVT) theory which contains the coupling of the scalar field to the double dual Riemann tensor and the field strength (i.e. $L^{\mu\nu\alpha\beta}F_{\mu\nu}F_{\alpha\beta}$).
We described this model as the $f_{4}$ model.
First, we investigated the stability of the test scalar field around a RN BH with the trivial scalar profile, and showed that the coupling induces the tachyonic instability in the vicinity of the BH event horizon if the BH charge or the coupling is sufficiently large and $H''(0)>0$.
The parameter region in which the scalarization is induced is depicted in Fig.\ref{graph:RN_instability_bdry_Sdeformation}.
\par
Secondly, using the test field approximation, the bound state of the scalar field around a RN BH was constructed.
To construct the profile, we consider the quadratic and the exponential models.
The quadratic model is the simplest model in which the scalarization can be realized.
The exponential model is a nonlinear example model.
Performing the stability analysis, we showed that the bound state of the quadratic model is unstable.
On the other hand, we also showed that the nonlinear effect of the scalar field in the exponential model can stabilize the bound state.
\par
Finally we constructed the scalarized BH solutions.
The solutions are characterized by the BH mass, electric charge, and the scalar charge.
The charge-to-mass ratio $Q/M$ depends on the charge, and there is the maximum value of $Q/M$.
It can be larger than unity.
Thus, scalarized BHs can be overcharged.
\par
The $f_{4}$ model has different properties from the previous models of scalarization\cite{Herdeiro:2018wub}.
In the previous model, the origin of scalarization is the nontrivial profile of $F^{\mu\nu}F_{\mu\nu}$.
So, scalarization can be induced in Minkowski spacetime with electric and/or magnetic fields\cite{Herdeiro:2018wub}.
In our model, the trigger of scalarization is the non-trivial profile of $L^{\mu\nu\alpha\beta}F_{\mu\nu}F_{\alpha\beta}$.
The term vanishes in Minkowski spacetime.
For scalarization of the $f_{4}$ model, the curved space-time is crucial.
\par
In this paper, we consider scalarization of the $f_{4}$ model around RN BHs, and showed that scalarization is induced.
In an astrophysical setup, scalarization may realize under the Kerr BH with the external magnetic field, which may be approximated by Wald's solution\cite{Wald:1974np}.
Therefore, it is expected that the difference between the $f_{4}$ model and Einstein-Maxwell theory with minimal coupling may appear in such a setup.
This task is beyond the scope of this paper.
\acknowledgments
T. I. acknowledges financial support provided under the European Union's H2020 ERC Consolidator Grant "Matter and strong-field gravity: New frontiers in Einstein's theory" grant agreement no. MaGRaTh-646597, and under the H2020-MSCA-RISE-2015 Grant No. StronGrHEP-690904.
M.M.~was supported by the research grant under the Decree-Law 57/2016 of August 29 (Portugal) through the Funda\c{c}\~{a}o para a Ci\^encia e a Tecnologia.
\appendix
\section{Scalar-vector-tensor theories with $U(1)$ symmetry}\label{sec:Scalar-vector-tensor theories with U(1) symmetry}
In this appendix, we summarize the most general SVT theory with $U(1)$ symmetry whose equations of motion are of the second order\cite{Heisenberg:2018acv}.
The scalar, vector and tensor fields are denoted by $\Phi$, $A_{\mu}$, and $g_{\mu\nu}$, respectively.
\par
The action of the theory can be expressed in the following form
\begin{eqnarray}
S[g,\Phi,A]&=&\int d^{4}x\sqrt{-g}\left(
\sum_{i=3}^{5}\mathcal{L}_{\rm ST}^{i}+\sum_{i=2}^{4}\mathcal{L}^{i}_{\rm SVT}
\right),\label{action SVT}
\end{eqnarray}
where $\mathcal{L}^{i}_{\rm ST} (i=3,4,5)$ are the Lagrangians of the ST sector, and $\mathcal{L}_{\rm SVT}^{i} (i=2,3,4)$ are the Lagrangians of the SVT sector.
The action of the ST part corresponds to the Horndeski theory\cite{Horndeski:1974wa,Kobayashi:2011nu,Kobayashi:2019hrl}, which is given as
\begin{widetext}
\begin{eqnarray}
\mathcal{L}_{\rm ST}^{3}&=&G_{3}(\Phi,X)\Phi_{\mu}^{~\mu},\\
\mathcal{L}_{\rm ST}^{4}&=&G_{4}(\Phi,X)R+G_{4,X}(\Phi,X)\left(
(\Phi_{\mu}^{~\mu})^{2}-\Phi_{\mu}^{~\nu}\Phi_{\nu}^{~\mu}
\right),\\
\mathcal{L}_{\rm ST}^{5}&=&G_{5}(\Phi,X)G_{\mu\nu}\Phi^{\mu\nu}-\frac{G_{5,X}(\Phi,X)}{6}\left(
(\Phi_{\mu}^{~\mu})^{3}
-3(\Phi_{\mu}^{~\mu})(\Phi_{\nu}^{~\rho}\Phi_{\rho}^{~\nu})
+2\Phi_{\mu}^{~\nu}\Phi_{\nu}^{~\rho}\Phi_{\rho}^{~\mu}
\right),
\end{eqnarray}
\end{widetext}
where $G_{i=3,4,5}(\Phi,X)$ are arbitrary functions of $\Phi$ and $X$, and $G_{i, X}(\Phi,X)=\partial_{X}G_{i}(\Phi,X)$.
\footnote{
The $G_{2}$ term of Horndeski theory is included in the $f_{2}$ term of scalar-vector-tensor sector.
}
Here, $\Phi_{\alpha\beta}=\nabla_{\alpha}\nabla_{\beta}\Phi$ is the 2nd covariant derivative of $\Phi$, and $X=-\frac{1}{2}g^{\mu\nu}\nabla_{\mu}\Phi\nabla_{\nu}\Phi$ is the canonical kinetic term of $\Phi$. 
$R$, and $G_{\mu\nu}$ are the Ricci scalar and the Einstein tensor.
\par
The SVT parts of the theory are given as
\begin{eqnarray}
\mathcal{L}_{\rm SVT}^{2}&=&f_{2}(\Phi,X,F,\tilde{F},Y),\\
\mathcal{L}_{\rm SVT}^{3}&=&\mathcal{M}_{3}^{\mu\nu}\Phi_{\mu\nu},\\
\mathcal{L}_{\rm SVT}^{4}&=&\mathcal{M}^{\mu\nu\alpha\beta}_{4}\Phi_{\mu\alpha}\Phi_{\nu\beta}
+f_{4}(\Phi,X)L^{\mu\nu\alpha\beta}F_{\mu\nu}F_{\alpha\beta},\nonumber\\
\end{eqnarray}
where $F$, $\tilde{F}$, and $Y$ are defined as
\begin{eqnarray}
F&=&F^{\mu\nu}F_{\mu\nu},\\
\tilde{F}&=&\tilde{F}^{\mu\nu}F_{\mu\nu},\\
Y&=&\nabla_{\mu}\Phi\nabla_{\nu}\Phi F^{\mu\alpha}F^{\nu}_{~\alpha},
\end{eqnarray}
and $\mathcal{M}_{3}^{\mu\nu}$, $\mathcal{M}_{4}^{\mu\nu\alpha\beta}$, and $L^{\mu\nu\alpha\beta}$ are given as
\begin{eqnarray}
\mathcal{M}_{3}^{\mu\nu}&=&\left(
f_{3}(\Phi,X)g_{\rho\sigma}+\tilde{f}_{3}(\Phi,X)\nabla_{\rho}\Phi\nabla_{\sigma}\Phi
\right)\tilde{F}^{\mu\rho}\tilde{F}^{\nu\sigma},\nonumber\\
\\
\mathcal{M}_{4}^{\mu\nu\alpha\beta}&=&
\left(
\frac{1}{2}f_{4,X}(\Phi,X)
+\tilde{f}_{4}(\Phi)
\right)
\tilde{F}^{\mu\nu}\tilde{F}^{\alpha\beta},\\
L^{\mu\nu\alpha\beta}&=&\frac{1}{4}\epsilon^{\mu\nu\rho\sigma}\epsilon^{\alpha\beta\gamma\delta}R_{\rho\sigma\gamma\delta}.
\label{eq:def dual Riemann}
\end{eqnarray}
Here, $R_{\mu\nu\alpha\beta}$ is the Riemann tensor,
$\epsilon^{\alpha\beta\gamma\delta}=\frac{1}{\sqrt{-g}}E^{\alpha\beta\gamma\delta}$ is the covariant anti-symmetric tensor, and $E^{\alpha\beta\gamma\delta}$ is the Levi-Civita tensor with $E^{0123}=1$.
$F_{\mu\nu}$ is the field strength (i.e. $F_{\mu\nu}=\nabla_{\mu}A_{\nu}-\nabla_{\nu}A_{\mu}$), and $\tilde{F}^{\mu\nu}=\frac{1}{2}\epsilon^{\mu\nu\lambda\rho}F_{\lambda\rho}$ is the dual of the field strength.
$f_{2}(\Phi,X,F,\tilde{F},Y)$ is an arbitrary function of $\Phi$, $X$, $F$, $\tilde{F}$, and $Y$.
$f_{3}(\Phi,X)$, $\tilde{f}_{3}(\Phi,X)$, and $f_{4}(\Phi,X)$ are arbitrary functions of $\Phi$ and $X$, respectively.
$\tilde{f}_{4}(\Phi)$ is an arbitrary function of $\Phi$.
$f_{4,X}(\Phi,X)$ is the derivative of $f_{4}(\Phi,X)$ with respect to $X$.
\par
The action Eq.\eqref{action SVT} gives a framework of the most general SVT theory with 2nd order equations of motion which has $U(1)$ symmetry.
This theory includes several classes of theory which has the scalar, vector, and tensor fields without an Ostrogradsky ghost.
One of the important subclasses is the theory which has the shift symmetry:
\begin{eqnarray}
\Phi\to\Phi + c.
\label{shift symmetry}
\end{eqnarray}
The theory with the shift symmetry is given by eliminating the $\Phi$ dependence as follows:
\begin{eqnarray}
G_{i=3,4,5}(\Phi,X)&\to&G_{i=3,4,5}(X),\\
f_{2}(\Phi,X,F,\tilde{F},Y)&\to&f_{2}(X,F,\tilde{F},Y),\\
f_{i=3,4}(\Phi,X)&\to&f_{i=3,4}(X),\\
\tilde{f}_{3}(\Phi,X)&\to&\tilde{f}_{3}(X),\\
\tilde{f}_{4}(\Phi)&\to&{\rm constant.}
\end{eqnarray}
By virtue of the shift symmetry, the equation of motion of the scalar field can be written as divergence form:
\begin{eqnarray}
\nabla_{\mu}J_{\Phi}^{\mu}=0,
\label{conservation of current}
\end{eqnarray}
where $J^{\mu}_{\Phi}$ is the Noether current associated with the shift symmetry.
\section{No hair theorem for shift symmetric SVT theory with $U(1)$ symmetry}\label{Sec:No hair theorem for subclass with shift symmetry}
In this section, we only consider the theory which possesses the shift symmetry Eq.\eqref{shift symmetry}, and discuss an extension of the BH no-hair theorem of the shift symmetric Horndeski theory\cite{Hui:2012qt} to the subclass of the SVT theory with the shift symmetry.
\par
We prove the no-hair theorem of the shift symmetric SVT theory without a cosmological constant under the following assumptions:
\begin{enumerate}
\item The spacetime is spherically symmetric and static spacetime with the asymptotic flatness.
\item $F$, $\tilde{F}$ and the gradient of the scalar field vanish in the asymptotic region.
\item The scalar and vector fields have the same symmetries with the metric.
\item The norm of the Noether current $J_{\Phi\mu}J^{\mu}_{\Phi}$ is finite on and outside the BH horizon.
\item The theory has the canonical kinetic term of the scalar field $X\subset f_{2}(X,F,\tilde{F},Y)$.
\item The functions $G_{i=3,4,5}(X)$, $f_{i=3,4}(X)$ and $\tilde{f}_{3}(X)$ are analytic at $X=0$.
\item $f_{2}(X,F,\tilde{F},Y)$ is analytic at $X=0$, $Y=0$, $F=0$, and $\tilde{F}=0$.
\item $f_{3}(X)$ vanishes at $X=0$.
\end{enumerate}
Since we assume that the action does not have a cosmological constant,  $f_{2}(0,0,0,0,0)=0$, and $f_{2}$ vanishes at the infinity.
The no-hair theorem states that under the above conditions, BHs can not have a nontrivial profile of the scalar field and are solutions of the following vector-tensor theory:
\begin{eqnarray}
S_{\rm VT}&=&\int d^{4}x\sqrt{-g}\left(
G_{4}(0,0)R+
f_{2}(0,0,F,\tilde{F},0)\right.\nonumber\\
&&\left.+f_{4}(0,0)L^{\mu\nu\alpha\beta}F_{\mu\nu}F_{\alpha\beta}
\right).
\end{eqnarray}
This theory contains RN BHs, Born-Infeld BHs\cite{Rasheed:1997ns}, and BH solutions of the generalized Einstein-Maxwell theory derived by Horndeski\cite{Horndeski:1978ca}.
The assumptions 1-7 are simple extensions from the assumptions of the no hair theorem of the shift symmetric Horndeski theory,
and the condition 8 is an additional one.
\par
To prove the no hair theorem, we focus on the general property of the current $J_{\Phi}^{\mu}$.
In this section, we use the following metric ansatz:
\begin{eqnarray}
ds^{2}&=&-f(r)dt^{2}+\frac{1}{f(r)}dr^{2}+g^{2}(r)r^{2}(d\theta^{2}+\sin^{2}\theta d\phi^{2}),\nonumber\\
\end{eqnarray}
where $f(r)$ and $g(r)$ are functions of $r$.
Without loss of generality, we assume $g(r)>0$.
The location $r_{\rm H}$ of the BH horizon is dictated by $f(r_{\rm H})=0$.
\par
Following the same discussion as the shift symmetric Horndeski theory \cite{Hui:2012qt},
the assumptions 1 and 3 imply that the nontrivial component of the current $J^{\mu}_{\Phi}$ is radial direction $J_{\Phi}^{r}$,
and, the assumption 4 implies that 
\begin{eqnarray}
J_{\Phi}^{r}=0 \label{eq:Jr vanish}
\end{eqnarray}
on and outside of the BH horizon.
For further discussion, we focus on the expression of the current $J_{\Phi}^{r}$.
The expression of $J_{\Phi}^{r}$ is expressed as
\begin{eqnarray}
J_{\Phi}^{r}&=&C_{f2Y}f_{2Y}(X,F,\tilde{F},Y)
+C_{f2X}f_{2X}(X,F,\tilde{F},Y)\nonumber\\
&&+C_{f3}f_{3}(X)
+C_{f3X}f_{3X}(X)\nonumber\\
&&+C_{f4X}f_{4X}(X)
+C_{f4XX}f_{4XX}(X)\nonumber\\
&&+C_{G3X}G_{3X}(X)
+C_{G4X}G_{4X}(X)
+C_{G4XX}G_{4XX}(X)\nonumber\\
&&+C_{G5X}G_{5X}(X)
+C_{G5XX}G_{5XX}(X),
\label{eq:shift symmetric current}
\end{eqnarray}
where $X=-\frac{1}{2}f(r)\Phi^{\prime 2}$, $\tilde{F}=0$, $F=-2A_{t}^{\prime 2}$, and $Y=-f(r)A_{t}^{\prime 2}\Phi^{\prime 2}$.
The coefficients $C_{f2Y}$, $C_{f2X}$, $C_{f3}$, $C_{f3X}$, $C_{f4X}$, $C_{f4XX}$, $C_{G3X}$, $C_{G4X}$, $G_{4XX}$, $G_{5X}$, and $C_{G5XX}$
are expressed as
\begin{widetext}
\begin{eqnarray}
C_{f2Y}&=&-F(r) f(r) \Phi '(r),\\
C_{f2X}&=&f(r) \Phi '(r),\\
C_{f3}&=&\frac{F(r) f(r) \left(r g'(r)+g(r)\right)}{r g(r)},\\
C_{f3X}&=&\frac{2 F(r) X(r) f(r) \left(r g'(r)+g(r)\right)}{r g(r)},\\
C_{f4X}&=&\frac{F(r) f(r) \Phi '(r) \left(3 f(r) \left(r g'(r)+g(r)\right)^2-2\right)}{r^2 g(r)^2},\\
C_{f4XX}&=&\frac{F(r) X(r) f(r)^2 \left(r g'(r)+g(r)\right)^2 \Phi '(r)}{r^2 g(r)^2},\\
C_{G3X}&=&X(r) \left(-f'(r)-\frac{4 f(r) \left(r g'(r)+g(r)\right)}{r g(r)}\right),\\
C_{G4X}&=&-\frac{2 f(r) \Phi '(r) \left(r^2 g(r) f'(r) g'(r)+r g(r)^2 f'(r)+f(r) \left(r g'(r)+g(r)\right)^2-1\right)}{r^2 g(r)^2},\\
C_{G4XX}&=&-\frac{4 X(r) f(r) \left(r g'(r)+g(r)\right) \Phi '(r) \left(r g(r) f'(r)+f(r) \left(r g'(r)+g(r)\right)\right)}{r^2 g(r)^2},\\
C_{G5X}&=&-\frac{X(r) f'(r) \left(3 f(r) \left(r g'(r)+g(r)\right)^2-1\right)}{r^2 g(r)^2},\\
C_{G5XX}&=&-\frac{2 X(r)^2 f(r) f'(r) \left(r g'(r)+g(r)\right)^2}{r^2 g(r)^2}.
\end{eqnarray}
\end{widetext}
Since the coefficient $C_{f3}$ does not vanish for $\Phi'=0$, 
the condition in which the solution with a trivial scalar field exists is $f_{3}(0)=0$, which is the assumption 8.
The expression of $X$, $F$, and $\tilde{F}$ ensures that these scalar functions are finite on and outside of BH horizon.
\par
We focus on the assumptions 6 and 7.
The assumptions imply that the functions $G_{i=3,4,5}(X)$, $f_{2}(X,F,\tilde{F},Y)$, $f_{i=3,4}(X)$, and $\tilde{f}_{3}(X)$ can be expressed in terms of Taylor series:
\begin{eqnarray}
G_{i=3,4,5}(X)&=&\sum_{N\geq0}\frac{G_{i=3,4,5}^{(N)}}{N!}X^{N},\\
f_{2}(X,F,\tilde{F},Y)&=&\sum_{N,M\geq0}\frac{f_{2}^{(N,M)}(F,\tilde{F})}{N!M!}X^{N}Y^{M},\\
f_{i=3,4}(X)&=&\sum_{N\geq0}\frac{f_{i=3,4}^{(N)}}{N!}X^{N},\\
\tilde{f}_{3}(X)&=&\sum_{N\geq0}\frac{\tilde{f}^{(N)}_{3}}{N!}X^{N},
\end{eqnarray}
where $f^{(0)}_{3}=0$.
Let us focus on the asymptotic region.
In the asymptotic region, the metric functions behave as $f(r),g(r)=1+\mathcal{O}(\frac{1}{r})$, and $F$,$\tilde{F}$ decay as some power of $\frac{1}{r}$.
Therefore, the form of the current in the asymptotic region is given as
\begin{eqnarray}
J^{r}\simeq \alpha \Phi^{\prime},
\end{eqnarray}
where 
\begin{eqnarray}
\alpha=f_{2}^{(1,0)}(0,0).
\end{eqnarray}
The assumption 5 implies that $\alpha$ does not vanish, and $\Phi'$ must be zero in the asymptotic region.
From Eq.\eqref{eq:shift symmetric current}, the current behaves as
\begin{eqnarray}
J^{r}=S(\Phi'(r),F(r),f(r),g(r))\Phi'(r),
\end{eqnarray}
where $S$ is a nonlinear function of $\Phi'(r)$,$F(r)$,$f(r)$, and $g(r)$, and approaches to $\alpha$ at infinity.
Moving inward from the infinity, $S$ varies continuously, and it implies that 
the solution of $J^{r}=0$ is $\Phi'=0$ everywhere outside the horizon.
This is the no hair theorem of the shift symmetric SVT theory with $U(1)$ symmetry.
\section{Validity of the test field analysis.}\label{appendix:Validity of the test field analysis}
In this appendix, the validity of the test field analysis used in Sec.\ref{Sec. Test field analysis} is discussed.
\par
The analysis can be valid under the limit of small $G$ in which the Einstein-Maxwell theory is the leading order and the effect of the scalar field is the next order.
To discuss the limit explicitly, let us introduce $\tilde{A}_{\mu}$ and $\tilde{H}(\Phi)$ which are defined as
\begin{eqnarray}
A_{\mu}&\equiv&\frac{1}{\sqrt{G}}\bar{A}_{\mu}.\\
H(\Phi)&\equiv&G\bar{H}(\Phi).
\end{eqnarray}
Using these variables, the action is expressed as
\begin{eqnarray}
S&=&\int d^{4}x\sqrt{-g}
\left[
\frac{1}{G}
\left(
\frac{1}{16\pi}R
-\frac{1}{4}\bar{F}^{\mu\nu}\bar{F}_{\mu\nu}\right)
-\frac{1}{2}\nabla^{\mu}\Phi\nabla_{\mu}\Phi
\right.
\nonumber\\
&&\left.
+\bar{H}(\Phi)L^{\mu\nu\alpha\beta}\bar{F}_{\mu\nu}\bar{F}_{\alpha\beta}
\right],
\end{eqnarray}
where $\bar{F}_{\mu\nu}=\nabla_{\mu}\bar{A}_{\nu}-\nabla_{\nu}\bar{A}_{\mu}$.
Taking the limit of small G with fixing the metric and $\bar{A}_{\mu}$, the leading order action becomes the Einstein-Maxwell theory, and the 1st order equation is the Klein-Gordon equation with the coupling.
This limit corresponds to the test field analysis in Sec.\ref{Sec. Test field analysis}.




\begin{thebibliography}{99}
\bibitem{Bekenstein:1971hc} 
  J.~D.~Bekenstein,
  Phys.\ Rev.\ D {\bf 5}, 1239 (1972).
  doi:10.1103/PhysRevD.5.1239

\bibitem{Bekenstein:1995un} 
  J.~D.~Bekenstein,
  Phys.\ Rev.\ D {\bf 51}, no. 12, R6608 (1995).
  doi:10.1103/PhysRevD.51.R6608
  
\bibitem{Herdeiro:2014goa} 
  C.~A.~R.~Herdeiro and E.~Radu,
  Phys.\ Rev.\ Lett.\  {\bf 112}, 221101 (2014)
  doi:10.1103/PhysRevLett.112.221101
  [arXiv:1403.2757 [gr-qc]].
  
\bibitem{Herdeiro:2016tmi} 
  C.~Herdeiro, E.~Radu and H.~Rúnarsson,
  Class.\ Quant.\ Grav.\  {\bf 33}, no. 15, 154001 (2016)
  doi:10.1088/0264-9381/33/15/154001
  [arXiv:1603.02687 [gr-qc]].
  
\bibitem{Abbott:2016blz} 
  B.~P.~Abbott {\it et al.} [LIGO Scientific and Virgo Collaborations],
  Phys.\ Rev.\ Lett.\  {\bf 116}, no. 6, 061102 (2016)
  doi:10.1103/PhysRevLett.116.061102
  [arXiv:1602.03837 [gr-qc]].

\bibitem{Akiyama:2019cqa} 
  K.~Akiyama {\it et al.} [Event Horizon Telescope Collaboration],
  Astrophys.\ J.\  {\bf 875}, no. 1, L1 (2019)
  doi:10.3847/2041-8213/ab0ec7
  [arXiv:1906.11238 [astro-ph.GA]].
  
\bibitem{Hui:2012jb} 
  L.~Hui and A.~Nicolis,
  Phys.\ Rev.\ Lett.\  {\bf 109}, 051304 (2012)
  doi:10.1103/PhysRevLett.109.051304
  [arXiv:1201.1508 [astro-ph.CO]].
  
\bibitem{Heisenberg:2018acv} 
  L.~Heisenberg,
  JCAP {\bf 1810}, no. 10, 054 (2018)
  doi:10.1088/1475-7516/2018/10/054
  [arXiv:1801.01523 [gr-qc]].
  
\bibitem{Heisenberg:2018vsk} 
  L.~Heisenberg,
  Phys.\ Rept.\  {\bf 796}, 1 (2019)
  doi:10.1016/j.physrep.2018.11.006
  [arXiv:1807.01725 [gr-qc]].
  
\bibitem{Horndeski:1974wa} 
  G.~W.~Horndeski,
  Int.\ J.\ Theor.\ Phys.\  {\bf 10}, 363 (1974).
  doi:10.1007/BF01807638
  
\bibitem{Kobayashi:2011nu} 
  T.~Kobayashi, M.~Yamaguchi and J.~Yokoyama,
  Prog.\ Theor.\ Phys.\  {\bf 126}, 511 (2011)
  doi:10.1143/PTP.126.511
  [arXiv:1105.5723 [hep-th]].
  
  
\bibitem{Kobayashi:2019hrl} 
  T.~Kobayashi,
  Rept.\ Prog.\ Phys.\  {\bf 82}, no. 8, 086901 (2019)
  doi:10.1088/1361-6633/ab2429
  [arXiv:1901.07183 [gr-qc]].
  
\bibitem{Woodard:2015zca} 
  R.~P.~Woodard,
  Scholarpedia {\bf 10}, no. 8, 32243 (2015)
  doi:10.4249/scholarpedia.32243
  [arXiv:1506.02210 [hep-th]].
    
\bibitem{Hui:2012qt} 
  L.~Hui and A.~Nicolis,
  Phys.\ Rev.\ Lett.\  {\bf 110}, 241104 (2013)
  doi:10.1103/PhysRevLett.110.241104
  [arXiv:1202.1296 [hep-th]].
  
\bibitem{Babichev:2016rlq} 
  E.~Babichev, C.~Charmousis and A.~Lehebel,
  Class.\ Quant.\ Grav.\  {\bf 33}, no. 15, 154002 (2016)
  doi:10.1088/0264-9381/33/15/154002
  [arXiv:1604.06402 [gr-qc]].
  
\bibitem{Heisenberg:2018vti} 
  L.~Heisenberg and S.~Tsujikawa,
  Phys.\ Lett.\ B {\bf 780}, 638 (2018)
  doi:10.1016/j.physletb.2018.03.059
  [arXiv:1802.07035 [gr-qc]].
  
\bibitem{Damour:1993hw} 
  T.~Damour and G.~Esposito-Farese,
  Phys.\ Rev.\ Lett.\  {\bf 70}, 2220 (1993).
  doi:10.1103/PhysRevLett.70.2220

\bibitem{Damour:1996ke} 
  T.~Damour and G.~Esposito-Farese,
  Phys.\ Rev.\ D {\bf 54}, 1474 (1996)
  doi:10.1103/PhysRevD.54.1474
  [gr-qc/9602056].
  
\bibitem{Harada:1998ge} 
  T.~Harada,
  Phys.\ Rev.\ D {\bf 57}, 4802 (1998)
  doi:10.1103/PhysRevD.57.4802
  [gr-qc/9801049].

\bibitem{Freire:2012mg} 
  P.~C.~C.~Freire {\it et al.},
  Mon.\ Not.\ Roy.\ Astron.\ Soc.\  {\bf 423}, 3328 (2012)
  doi:10.1111/j.1365-2966.2012.21253.x
  [arXiv:1205.1450 [astro-ph.GA]].
  
\bibitem{Berti:2015itd} 
  E.~Berti {\it et al.},
  Class.\ Quant.\ Grav.\  {\bf 32}, 243001 (2015)
  doi:10.1088/0264-9381/32/24/243001
  [arXiv:1501.07274 [gr-qc]].
  
\bibitem{Silva:2017uqg} 
  H.~O.~Silva, J.~Sakstein, L.~Gualtieri, T.~P.~Sotiriou and E.~Berti,
  Phys.\ Rev.\ Lett.\  {\bf 120}, no. 13, 131104 (2018)
  doi:10.1103/PhysRevLett.120.131104
  [arXiv:1711.02080 [gr-qc]].

\bibitem{Antoniou:2017acq} 
  G.~Antoniou, A.~Bakopoulos and P.~Kanti,
  Phys.\ Rev.\ Lett.\  {\bf 120}, no. 13, 131102 (2018)
  doi:10.1103/PhysRevLett.120.131102
  [arXiv:1711.03390 [hep-th]].
  
\bibitem{Doneva:2017bvd} 
  D.~D.~Doneva and S.~S.~Yazadjiev,
  Phys.\ Rev.\ Lett.\  {\bf 120}, no. 13, 131103 (2018)
  doi:10.1103/PhysRevLett.120.131103
  [arXiv:1711.01187 [gr-qc]].
  
\bibitem{Antoniou:2017hxj} 
  G.~Antoniou, A.~Bakopoulos and P.~Kanti,
  Phys.\ Rev.\ D {\bf 97}, no. 8, 084037 (2018)
  doi:10.1103/PhysRevD.97.084037
  [arXiv:1711.07431 [hep-th]].

\bibitem{Minamitsuji:2018xde} 
  M.~Minamitsuji and T.~Ikeda,
  Phys.\ Rev.\ D {\bf 99}, no. 4, 044017 (2019)
  doi:10.1103/PhysRevD.99.044017
  [arXiv:1812.03551 [gr-qc]].

\bibitem{Silva:2018qhn} 
  H.~O.~Silva, C.~F.~B.~Macedo, T.~P.~Sotiriou, L.~Gualtieri, J.~Sakstein and E.~Berti,
  Phys.\ Rev.\ D {\bf 99}, no. 6, 064011 (2019)
  doi:10.1103/PhysRevD.99.064011
  [arXiv:1812.05590 [gr-qc]].
  
\bibitem{Cunha:2019dwb} 
  P.~V.~P.~Cunha, C.~A.~R.~Herdeiro and E.~Radu,
  Phys.\ Rev.\ Lett.\  {\bf 123}, no. 1, 011101 (2019)
  doi:10.1103/PhysRevLett.123.011101
  [arXiv:1904.09997 [gr-qc]].
  
\bibitem{Minamitsuji:2019iwp} 
  M.~Minamitsuji and T.~Ikeda,
  Phys.\ Rev.\ D {\bf 99}, no. 10, 104069 (2019)
  doi:10.1103/PhysRevD.99.104069
  [arXiv:1904.06572 [gr-qc]].
  
\bibitem{Anson:2019uto} 
  T.~Anson, E.~Babichev, C.~Charmousis and S.~Ramazanov,
  JCAP {\bf 1906}, no. 06, 023 (2019)
  doi:10.1088/1475-7516/2019/06/023
  [arXiv:1903.02399 [gr-qc]].
  
\bibitem{Franchini:2019npi} 
  N.~Franchini and T.~P.~Sotiriou,
  arXiv:1903.05427 [gr-qc].
  
\bibitem{Herdeiro:2018wub} 
  C.~A.~R.~Herdeiro, E.~Radu, N.~Sanchis-Gual and J.~A.~Font,
  Phys.\ Rev.\ Lett.\  {\bf 121}, no. 10, 101102 (2018)
  doi:10.1103/PhysRevLett.121.101102
  [arXiv:1806.05190 [gr-qc]].
  
\bibitem{Fernandes:2019rez} 
  P.~G.~S.~Fernandes, C.~A.~R.~Herdeiro, A.~M.~Pombo, E.~Radu and N.~Sanchis-Gual,
  Class.\ Quant.\ Grav.\  {\bf 36}, no. 13, 134002 (2019)
  doi:10.1088/1361-6382/ab23a1
  [arXiv:1902.05079 [gr-qc]].
  
\bibitem{Fernandes:2019kmh} 
  P.~G.~S.~Fernandes, C.~A.~R.~Herdeiro, A.~M.~Pombo, E.~Radu and N.~Sanchis-Gual,
  arXiv:1908.00037 [gr-qc].
  
\bibitem{Stefanov:2007eq} 
  I.~Z.~Stefanov, S.~S.~Yazadjiev and M.~D.~Todorov,
  Mod.\ Phys.\ Lett.\ A {\bf 23}, 2915 (2008)
  doi:10.1142/S0217732308028351
  [arXiv:0708.4141 [gr-qc]].
  
\bibitem{Kimura:2017uor} 
  M.~Kimura,
  Class.\ Quant.\ Grav.\  {\bf 34}, no. 23, 235007 (2017)
  doi:10.1088/1361-6382/aa903f
  [arXiv:1706.01447 [gr-qc]].
 
  
\bibitem{Wald:1974np} 
  R.~M.~Wald,
  Phys.\ Rev.\ D {\bf 10}, 1680 (1974).
  doi:10.1103/PhysRevD.10.1680
  
\bibitem{Rasheed:1997ns} 
  D.~A.~Rasheed,
  hep-th/9702087.
  
\bibitem{Horndeski:1978ca} 
  G.~W.~Horndeski,
  Phys.\ Rev.\ D {\bf 17}, 391 (1978).
  doi:10.1103/PhysRevD.17.391
  
%
%
  
  
  
\end{thebibliography}
\end{document}